\documentclass[prb,twocolumn,showpacs,superscriptaddress,groupedaddress]{revtex4}
\usepackage{graphicx}
\usepackage{amsmath}
\usepackage{amssymb,color}

\def\be{\begin{equation}}
\def\ee{\end{equation}}
\def\ba{\begin{eqnarray}}
\def\ea{\end{eqnarray}}
\def\beq{\begin{equation}}
\def\eeq{\end{equation}}
\def\bea{\begin{eqnarray}}
\def\eea{\end{eqnarray}}
\begin{document}

\title{Superconductivity vs bound state formation in a two-band superconductor with small Fermi energy -- applications to Fe-pnictides/chalcogenides and doped SrTiO$_3$ }

\author{Andrey V. Chubukov}
\affiliation{Department of Physics, University of Minnesota, Minneapolis, Minnesota 55455, USA}
\author{Ilya Eremin}
\affiliation{Institut fur Theoretische Physik III, Ruhr-Universitat Bochum, D-44801 Bochum, Germany}
\author{Dmitri V.~Efremov}
\affiliation{Leibniz-Institut fur Festkorper- und Werkstoffforschung, D-01069 Dresden, Germany}
\begin{abstract}
  We analyze the interplay between superconductivity and the formation of bound pairs of fermions (BCS-BEC crossover)
 in a 2D model of interacting fermions with small Fermi energy $E_F$ and weak attractive interaction, which extends to energies well above $E_F$.
 The 2D case is special because two-particle bound state forms at arbitrary weak interaction, and already at weak coupling one has to distinguish between bound state formation and superconductivity. We briefly review the situation in the one-band model and then consider two different two-band models -- one  with  one hole band and one electron band and another with two hole or two electron bands. In each case we obtain the bound state energy $2E_0$ for two fermions in a vacuum and solve the set of coupled equations for the pairing gaps and the chemical potentials to obtain the onset temperature of the pairing, $T_{ins}$ and the quasiparticle dispersion at $T=0$.  We  then compute the superfluid stiffness $\rho_s (T=0)$ and obtain the actual $T_c$.
 For definiteness, we set $E_F$ in one band to be near zero and consider different ratios of $E_0$ and $E_F$ in the other band. We show that, at $E_F \gg E_0$, the behavior of both two-band models is BCS-like in the sense that  $T_{c} \approx T_{ins} \ll E_F$ and $\Delta \sim T_{c}$. At $E_F < E_0$, the two models behave differently:  in the model with two hole/two electron bands, $T_{ins} \sim E_0/\log{\frac{E_0}{E_F}}$, $\Delta \sim (E_0 E_F )^{1/2}$,  and $T_c \sim E_F$, like in the one-band model. In between $T_{ins}$ and $T_c$ the system displays preformed pair behavior.
 In the model with one hole and one electron band, $T_c$ remains of order  $T_{ins}$, and both remain finite at $E_F =0$ and of order $E_0$.  The preformed pair behavior still does exist in this model because $T_c$ is  numerically smaller than $T_{ins}$.
  For both models  we re-express $T_{ins}$ in terms of the fully renormalized two-particle scattering amplitude by extending to two-band case the
 method pioneered by Gorkov and Melik-Barkhudarov back in 1961.  We apply our results for the model with a hole and an electron band to Fe-pnictides and Fe-chalcogenides in which  superconducting gap has been detected on the bands which do not cross the Fermi level, and to FeSe, in which  superconducting gap is comparable to the Fermi energy.   We apply the results for the model with two electron bands to Nb-doped SrTiO$_3$ and argue that our theory explains rapid increase of $T_c$ when both bands start crossing the Fermi level.
\end{abstract}
\date{\today}

\maketitle

\section{Introduction}
The discovery of superconductivity in Fe-pnictides and later in Fe-chalcogenides opened up several new directions in the study of non-phononic mechanisms of  electronic pairing in multi-band  correlated electron systems~\cite{kamihara,reviews}.
Two issues were brought about by recent angle-resolved photoemission and other experiments in Fe-based superconductors (FeSCs).  First, in recent experiments  on LiFe$_{1-x}$Co$_x$As, Miao et al. observed\cite{Miao2015}  a finite superconductive gap of $4-5$meV on the hole band, which is located below the Fermi level, with the top of the band at 4-8meV away from $E_F$. Moreover, the gap on this hole band is larger than the gaps on electron bands, which cross the Fermi level.   A similar observation has been reported for FeTe$_{0.6}$Se$_{0.4}$ \cite{Ozaki2014}, where superconductivity with the gap  $\Delta = 1.3$meV has been observed on an electron band which lies above the Fermi level, with the bottom of the band at 0.7 meV  away from $E_F$.
 Second, recent photoemission
  measurements have demonstrated that in almost all Fe-based supperconductors either electron or hole pockets are more tiny than previously thought and the corresponding dispersions either barely cross the Fermi level or are fully located below or above it~\cite{borise}.  The "extreme" case in this respect is FeSe.  In this material
  Fermi energies on {\it all} hole and electron pockets are small and are comparable to the magnitudes of the superconducting gaps  (the reported $E_F$ on different bands vary between 4 and 10 meV, while the gaps are 3-5meV~\cite{Terashima2014, Kasahara2014}).

The observation of a sizable superconducting gap on a band which does not cross the Fermi level was originally interpreted~\cite{Miao2015}  as the indication that the pairing in the FeSCs is a strong coupling phenomenon for which the pairing gap is not confined to the Fermi surface and develops at all momenta in the Brillouin zone. Later, however, the  experiments were re-interpreted~\cite{Chen2015} in a more conventional weak/moderate coupling scenario, as the consequence of  the fact that in FsSCs the pairing interaction primarily "hopes" a pair of fermions with momenta ${\bf k}$ and $-{\bf k}$ from one band to the other~\cite{Kuroki2008,Mazin2008}. In this situation, the gap on the band, which does not cross the Fermi level is determined by the density of states at $E_F$ of the band which does cross the Fermi level. A solution of the coupled set of BCS gap equations for fermions in the two bands  then shows that one crossing is sufficient to obtain BCS instability already at weak coupling\cite{Chen2015}.  This reasoning also naturally explains why the gap is larger
on a band which does not cross the Fermi level.

The observation of superconductivity with $\Delta \sim E_F$  brought FeSCs  into the orbit of long-standing discussion about the interplay between superconductivity and the formation of the bound pair of two fermions.
This issue has been discussed  in the condensed matter context~\cite{Miyake1983,Nozieres85,Mohit89,Ohashi,DeMelo1993,ioffe97,tobij2000,lara2001,DeMelo2006,Ohashi2003,lara2009,Innocenti2010,Bianconi2013, levchenko2011,lara2012,Guidini2014,Mohit2015,kagan}. and  also for optical lattices of ultracold atoms. \cite{s_1}. The phenomenon in which bound pairs of fermions form at a higher $T_{ins}$ and condense at a smaller $T_c$
is often termed  Bose-Einstein condensation (BEC) because the condensation of preformed pairs (i.e., the development of a macroscopic condensate) bears a direct analogy with BEC of bosons in a Bose gas.  When $\Delta$ and $T_c$ are much smaller than $E_F$,  bound pairs and true superconductivity develop at almost the same  temperature, i.e., $T_c \approx T_{ins}$.  However, when $E_F$ gets smaller, superconducting $T_c$ is generally smaller than the onset temperature for bound state formation.

\begin{figure}[t!]
\renewcommand{\baselinestretch}{1.0}
\includegraphics[angle=0,width=0.6\linewidth]{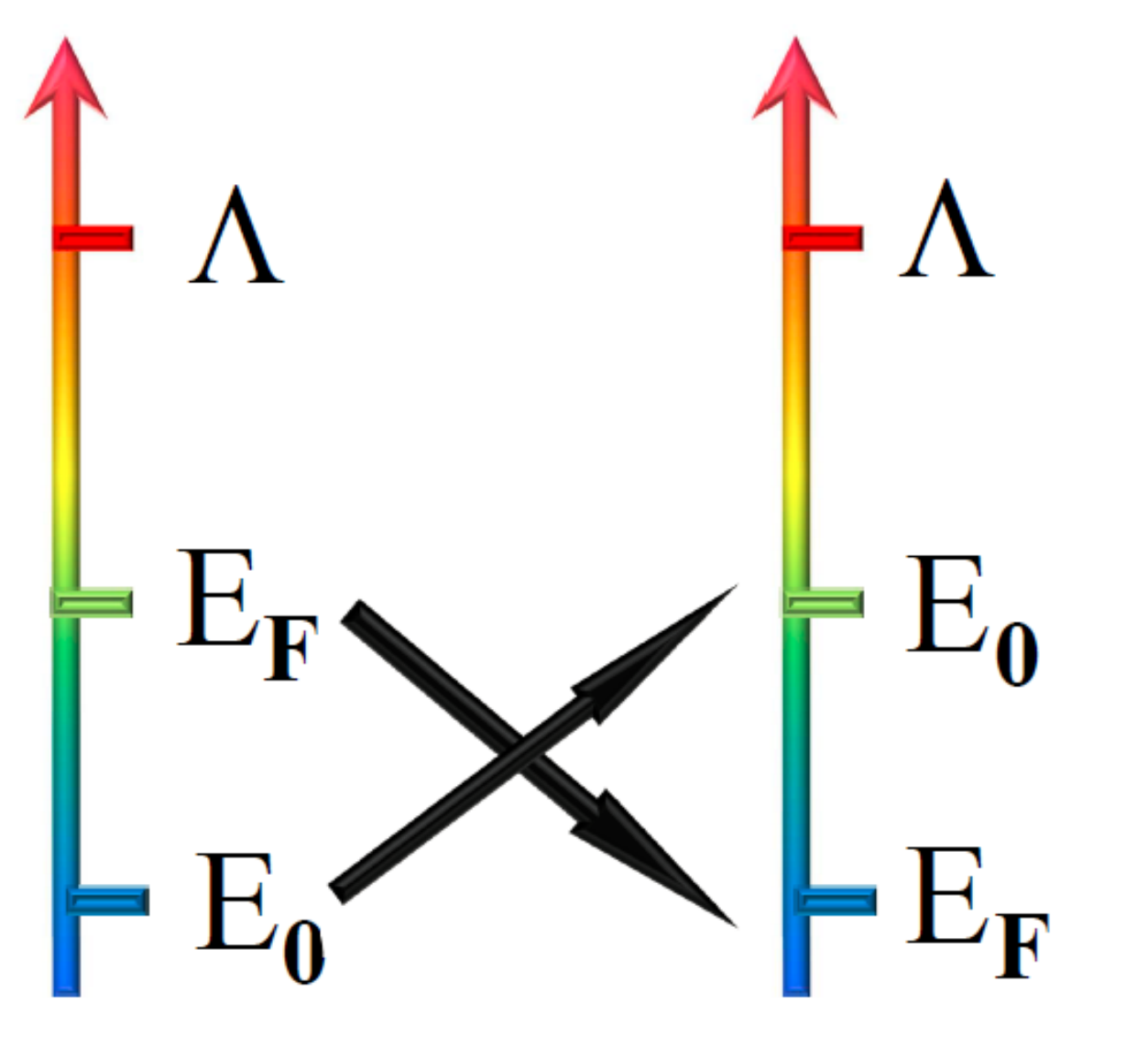}
\caption{Characteristic energy scales relevant to the interplay between the formation of bound pairs of fermions and true superconductivity in 2D Fermionic systems with weak attractive pairing interaction. $\Lambda$ is the upper energy cutoff, $E_F$ is the Fermi energy, and $E_0$ is the  energy of a bound state of two fermions in a vacuum, i.e., at $\mu =0$.  At weak coupling, $E_0 \ll \Lambda$.  We assume that $E_F$ is also small and can be tuned by doping to be either larger or smaller than $E_0$.
 We show that in one-band model and in two band model with two hole or two electron bands, the system displays BCS-like behavior at $E_F \gg E_0$ and BEC-like behavior at $E_F \ll E_0$.  In the latter case, bound pais develop at $T_{ins} \sim E_0/\log\frac{E_0}{E_F}$ but the true superconductivity with full phase coherence develops at $T_c \sim E_F$. In between $T_{ins}$ and $T_c$ the system displays preformed pair behavior and  the spectral function displays pseudogap behavior.
  In the two-band model with one hole and one electron pocket, $T_{ins}$ and $T_c$ also split when $E_F$ gets smaller than $E_0$, but both remain of order $E_0$
  even when $E_F$ vanishes.  Still, superconducting $T_c$ in this limit is several times smaller than $T_{ins}$, so there is a wide temperature range of preformed pair behavior.}%
\label{fig0}
\end{figure}

In the present communication, we discuss  superconductivity vs bound state formation in 2D systems with weak attractive interaction $U$ in the proper symmetry channel.
( $s^{+-}$ for the two-band model for FeSCs).  We consider the situation when $U$ remains energy independent  up to an energy $\Lambda$, which well exceeds $E_F$, see Fig.\ref{fig0}.
Elementary quantum mechanics shows that in  2D, two fermions with dispersion $k^2/(2m)$  form a bound state at arbitrary small attraction $U$, and the bound state
 energy is  $2E_0$, where $E_0 \sim \Lambda e^{-2/\lambda}$, and $\lambda = m U/(2\pi)$ is a dimensionless coupling.
  We  analyze the evolution of the system behavior and the interplay between $T_c$ and $T_{ins}$  by varying the ratio $E_F/E_0$ while keeping both $E_F$ and $E_0$ well below $\Lambda$.

We briefly review BCS-BEC crossover in the one-band  two-dimensional (2D) model and consider two different two-band models.

 The first model, which we apply to FeSCs, consists  of one hole and one electron band. We follow usual path and consider the case when the  dominant pairing interaction is weak inter-band pair hopping interaction $U>0$, in which case superconducting state has $s^{+-}$ symmetry.  We use the same computational procedure as in the studies of one-band model.~\cite{Miyake1983,Mohit89,DeMelo1993,DeMelo2006,Ohashi2003,Guidini2014,kagan} Namely, we  first obtain the bound state energy $2E_0$ for two fermions in a vacuum.  Then we consider the actual system with a non-zero density of carriers in one of the bands $n = 2 N_0 E_F$, where $N_0$ is a 2D density of states at low energies, and solve the set of coupled equations for the  chemical potential $\mu (T)$ and the pairing gaps  at a finite $T$.  The solution of the linearized  gap equations  yields  the onset temperature of the pairing $T_{ins}$. The solution of non-linear gap equations at $T=0$ yields the pairing gaps $\Delta_{h,e}$. We next use the values of $\Delta_{h,e}$ and $\mu$ at $T=0$ as inputs and compute superfluid stiffness $\rho_s (T=0)$.  For definiteness, we consider the case when the chemical potential at $T=0$ {\it in the would be normal state} is at the bottom of  the electron band but crosses the hole band i.e.,  the Fermi energy is zero for the electron band  but finite  for the hole band.

We present the results in the two limits when $E_F$ is either larger or smaller than $E_0$ and in the case when $E_F=E_0$. We show that in all cases the pairing gap develops on both bands and is larger on the electron band (the one which does not cross the Fermi level).  This agrees qualitatively but not quantitatively with the results obtained previously within the conventional BCS theory~\cite{Chen2015,phillips2015}, neglecting the renormalization of the chemical potential.
 We argue, however, that the renormalization of the chemical potential  by both a finite temperature and a finite gap is not a small perturbation
   when the bare chemical potential touches the bottom of the electron  band.

 We argue that the onset temperature of the pairing, $T_{ins}$ (the one obtained by solving the set of equations for the pairing gaps and the chemical potentials) evolves when $E_F/E_0$ changes and is of order $E^{1/3}_F E^{2/3}_0$ when $E_F \gg E_0$ and of order $E_0$ when $E_F \ll E_0$.  We further ague that superconducting $T_c$ is of order $T_{ins}$ for arbitrary $E_F/E_0$, but is numerically smaller than $T_{ins}$.  The numerical smallness implies that there exists a finite range of temperatures between $T_{ins}$ and $T_c$, where pairs are already formed but their phases are random and there is no superconductivity (the "preformed pairs" regime).

The emergence of the bound pairs at $T_{ins} > T_c$ and the existence of the preformed pairs regime is often associated with the crossover from BCS to BEC behavior. Such crossover has been studied in detail in the finite $T$ analysis of 3D one-band model (Ref. \cite{DeMelo1993}).  The temperatures $T_{ins}$ and $T_c$  were found to differ strongly at $E_F \ll E_0$: $T_{ins} \sim E_0/\log{\frac{E_0}{E_F}} \gg E_F$, while
 $T_c \sim E_F$, i.e., the ratio $T_c/T_{ins}$ vanishes at $E_F \to 0$. The behavior in the 2D case is quite similar (see below).
 In our two-band model the behavior is similar to the one-band model in that $T_{ins}$ becomes parametrically larger than $E_F$ at $E_F \ll E_0$, but differs in that  $T_c$ remains finite and of the same order as order $T_{ins}$ even at $E_F \to 0$.
  The reason, as we argue below, is that the development of the pairing gap below $T_{ins}$ reconstructs the fermionic dispersion and creates images, with opposite dispersion, of original hole and electron bands. This in turn gives rise to the shift of fermionic density  from the filled hole band and empty electron band into these image bands.   As a consequence, there appear new hole-like and electron-like bands with a finite density of carriers in each band, proportional to $T_{ins}$.   Superconducting $T_c$ scales with this density and is a fraction of $T_{ins}$. We show that preformed pair behavior still exists at $E_F \ll E_0$, but only because $T_c$, set by superconducting stiffness,  is numerically smaller than $T_{ins}$.

 We next consider the two-band model consisting of  two electron bands. We again assume that the dominant pairing interaction is inter-band pair hopping $U >0$ and  that the chemical potential is at or near the bottom of one of the bands, but crosses the dispersion of the other band at some finite $E_F$.  We show that,  at $E_F \gg E_0$, the behavior of this model is nearly identical to that in the model with a hole and an electron band. However, in the opposite limit $E_F << E_0$, the behavior of the model with two electron bands differs qualitatively from that of the model with a hole and an electron band and is quite similar to the behavior of
     the 2D one-band model in the BEC limit.   Namely, $T_{ins} \sim E_0/\log{\frac{E_0}{E_F}}$ and $T_c \sim E_F$, such that the ratio $T_c/T_{ins}$ vanishes at $E_F =0$.
      The reason is that for the two  bands with the same sign of dispersion, there are no free carriers at $E_F \to 0$, hence the pairing cannot create images of the original bands. Indeed, we show that in the model with two electron bands, the pairing gap $\Delta$, which is responsible for the Fermi surface reconstruction,
     scales  at $T=0$  as $\sqrt{E_0 E_F}$ and vanishes at $E_F =0$. The same behavior holds at $T=0$ in the 2D one-band model.~\cite{Miyake1983,Mohit89}

     Superconductivity in a system with two electron bands is realized experimentally in Nb-doped SrTiO$_3$ and, possibly, in heterostructures of LaAlO$_3$ and SrTiO$_3$~ (see Refs. \cite{Fernandes2013,Innocenti2010,Bianconi2013} and references therein).   The Fermi energy in one of the bands is finite already at zero doping and $E_F$ is likely much larger than $E_0$.
        The other electron band is above  the chemical potential at zero doping, but the chemical potential $\mu$  moves up with doping and enters this band once it
         exceeds the critical value $\mu^*$.   The data indicate~\cite{marel,joshua} that, when this happens,  $T_c$ rapidly increases.   To analyze this behavior we compute  $T_{c}$
          for  $\mu \neq \mu^*$. We show (see Fig. \ref{fig9}) that $T_c$ indeed increases when $\mu$ exceeds $\mu^*$, and
           the rate of the increase  is $(1-\mu^*/\mu) (E_F/E_0)^{2/3}$, i.e., it is enhanced by a large ratio of $E_F/E_0$.

\begin{widetext}

\begin{table}[t!]%

\begin{tabular}
[c]{|c|c|c|c|}\hline
& single-band  & 2 electron (hole) bands & 1 electron and 1 hole bands \\\hline
$E_F \gg E_0$ & $T_{ins} \sim \sqrt{E_F E_{0}}$, $\mu(T_{ins})\approx E_F$ & $T_{ins} \sim E^{1/3}_F E^{2/3}_0$, $\mu_1 (T_{ins})  \approx E_F$ & $T_{ins} \sim E^{1/3}_F E^{2/3}_0$, $\mu_e (T_{ins})  \approx -0.5 T_{ins}$ \\
& $T_c \approx T_{ins}$, $\Delta = 2 \sqrt{E_F E_0}$ & $\mu_2 (T_{ins}) \approx -0.5 T_{ins}$, $T_{c} \approx T_{ins}$ &  $\mu_h (T_{ins}) \approx E_F$, $T_c \approx T_{ins}$
\\
&   & $\Delta  \approx 1.78 E^{1/3}_F E^{2/3}_0$ & $\Delta  \approx 1.78 E^{1/3}_F E^{2/3}_0$,  \\
&   &   &  \\\hline
$E_F \ll E_0$ & $T_{ins} \approx E_0/\log{\frac{E_0}{E_F}}$, $\mu(T_{ins})\approx -E_0$ &  $T_{ins} \sim 4.5E_0/\log{\frac{E_0}{E_F}}$, $\mu_1 (T_{ins})  \approx -2.3 E_0$
 & $T_{ins} \sim 1.13 E_0\left(1+  0.22\frac{E_F}{E_0}\right)$,   \\
& $T_c \sim E_F/8  \ll T_{ins}$, $\Delta  = 2 \sqrt{E_F E_0}$ & $\mu_2 (T_{ins})  \approx -2.3 E_0$, $T_c \sim E_F/8\ll T_{ins}$,  & $\mu_h (T_{ins}) \approx 3 E_F/2$, $\mu_e (T_{ins}) \approx  - \frac{E_F}{2}$\\
&   & $\Delta\approx\sqrt{2E_F E_0}$ &  $T_c \sim 0.22 T_{ins}$, $\Delta=1.76 T_{ins}$ \\\hline
$E_F = E_0$ & $T_{ins} = 1.09 E_F$, $\mu (T_{ins}) = -0.09 E_F$ & $T_{ins} \sim 0.9 E_F$, $\mu_1  (T_{ins}) = 0.1 E_F$ & $T_{ins} = 1.35 E_F$, $\mu_e  (T_{ins}) = -0.35 E_F$\\
&  & $\mu_2 (T_{ins}) = -0.9 E_F$,  $\Delta  =  1.4 E_F$ & $\mu_h (T_{ins}) = 1.35 E_F$,  $T_c \sim 0.4 E_F$  \\
&   &  &  $\Delta = 2.4 E_F$ \\\hline
\end{tabular}
\caption{The summary of the results for the onset temperature of the pairing, $T_{ins}$, the actual superconducting transition temperature, $T_c$, the gap magnitude at $T=0$,  $\Delta$, and the chemical potentials, $\mu_i$,  for different ratios of the Fermi energy, $E_F$, and the bound state energy of two fermions in a vacuum, $E_0$.
  For the one band model, superconductivity is an ordinary $s-$wave. For the two-band models, superconducting state has $s^{\pm}$- symmetry (different signs of the gaps on the two bands). The gap magnitudes on different bands are approximately the same in the two-band models, up to small corrections, but nevertheless the gap on the band which
   doesn't cross the chemical potential at $T \geq T_{ins}$   is larger than that on the other band, which crosses the chemical potential.
   The chemical potentials satisfy $\mu_h+\mu_e=E_F$ for the two band model with a hole and an electron band, and $\mu_1-\mu_2=E_F$ for the model with two electron bands. }%
\label{table:drho_freq}%
\end{table}
\end{widetext}

Another goal of our work is to compare the analysis of BCS/BEC crossover with the approach put forward by Gorkov and Melik-Barkhudarov (GMB) back in 1961 (Refs.\cite{Gorkov61,extra_1}).  GMB considered a one-band model with attractive Hubbard interaction $U$ at weak coupling in D=3. They argued that superconductivity comes from fermions with energies not exceeding $E_F$, while all  contributions to the pairing susceptibility from fermions with higher energies can be absorbed into the renormalization of the original 4-fermion interaction into quantum-mechanical scattering amplitude.   GMB explicitly separated  the Cooper  logarithm (associated with the presence of a sharp Fermi surface at $E_F \neq 0$)  from the renormalization of the interaction into the scattering amplitude and obtained
 $T_c  =0.277 E_F e^{-\pi/(2|a|k_F)}$, where $a$ is the s-wave scattering length. They argued that this is the right formula for comparison with the experimental data because the scattering length  is the physically observable parameter, while the  interaction $U$ is not.

The GMB analysis does not include phase fluctuations, hence their instability temperature is the same as $T_{ins}$,  re-expressed in terms of scattering amplitude.  In the original GMB analysis (which we review in Sec. \ref{sec_2a} below) $|a|k_F$ is assumed to be small, and  no bound state develops.   We extend GMB analysis  to one-band and two-band models in 2D and will specifically consider the limit $E_F< E_0$.  In this limit,  the scattering amplitude diverges at the onset of bound state development at $T \sim E_0$ and changes sign at a smaller $T$. It is then  {\it a--prior}i unclear whether the onset temperature of the pairing, $T_{ins}$, can be expressed via the 2D scattering amplitude $a_2$ (dimensionless in 2D) with $E_F$ in the prefactor, particularly given that the ratio $T_{ins}/E_F$ tends to infinity when $E_F =0$.  We, however, show that GMB approach remains valid even when  $E_F< E_0$, and $T_{ins}$ can be explicitly expressed via the exact 2D scattering amplitude, with $E_F$ in the prefactor.

The paper is organized as follows.  In Sec.\ref{sec_2} we review one-band 2D Fermi system with small $E_F$. We reproduce earlier results~\cite{Miyake1983,Mohit89,kagan} for the onset temperature for the pairing, $T_{ins}$, the  pairing gap, the renormalized chemical potential, and the spin stiffness.
  We argue that superconducting $T_c$ scales with $E_F$ and vanishes when $E_F=0$.
 In Sec. \ref{sec_3}  we consider in detail the case of one hole and one electron bands, relevant to FeSCs.  We show that in this model both $T_{ins}$ and $T_c$ remain finite even when neither bands crosses the Fermi level. The superconducting $T_c$ is smaller than $T_{ins}$ in this case, but the smallness is only  numerical.
  In Sec. \ref{sec_3_n} we consider the case of two hole/two electron pockets and show that  that $T_c$ remains non-zero as long as one of the band crosses Fermi level but vanishes when $E_F=0$ for both bands.
 In Sec. \ref{sec_2a} we review GMB formalism and then apply  it first to 2D one-band model and then to 2D model with a hole and an electron band.  In both cases we show that the instability temperature $T_{GMB}$ is precisely $T_{ins}$, even when $E_F \to 0$.  We present our conclusions in Sec. \ref{sec_4}. Discussion of some technical details is moved into the Appendix.\\

 \section{One-band model}
 \label{sec_2}

\begin{widetext}

\begin{figure}[t!]
\renewcommand{\baselinestretch}{.8}
\includegraphics[angle=0,width=.8\linewidth]{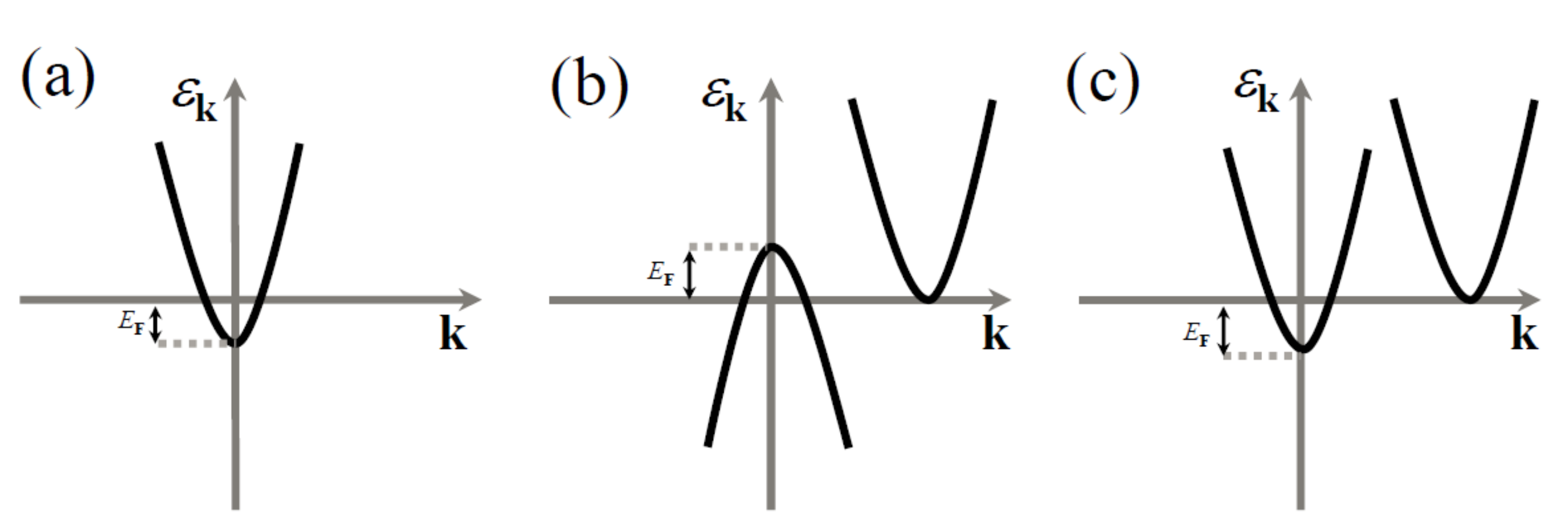}
\caption{The bare dispersion for the  models considered in the present manuscript: (a) a one-band model of 2D fermions with the parabolic dispersion and a positive bare chemical potential (i.e., a non-zero $E_F$);  (b) a two-band model with one hole and one electron band separated in the momentum space.  For definiteness, we set the bare chemical potential such that
  it touches the bottom of the electron band and crosses the hole band at a finite distance from its top;  (c) a two-band model
   with two electron bands separated in the momentum space. For definiteness we set the bare chemical potential to touch the bottom of one band and cross the other. }%
\label{fig1}
\end{figure}
\end{widetext}

To set the stage for the analysis of the two-band model we first review pairing and superconductivity in the 2D one-band model.
Consider a set of 2D fermions with the parabolic dispersion $\varepsilon_k = \frac{k^2}{2m}$ and chemical potential $\mu_0 = E_F$, see Fig.\ref{fig1}(a).
We assume that fermions get paired by a weak attractive pairing interaction $U(q, \Omega)$, which for simplicity we approximate as momentum and frequency independent $U$ up to upper momentum cutoff $q_{max}$ and corresponding frequency cutoff $\Lambda = q^2_{max}/(2m)$. For electron-phonon interaction $\Lambda$ is of the order of Debye frequency. The actual dispersion does not have to be parabolic, however at weak coupling energies relevant for the pairing are much smaller than $\Lambda$ and $\varepsilon_k = \frac{k^2}{2m}$ can be just viewed as the leading term in the expansion of the lattice dispersion in small momentum.

We introduce the dimensionless parameter
\beq
   \lambda = N_0 |U| = \frac{m |U|}{2\pi},
\eeq
where $N_0 = \frac{m}{2\pi}$ is the density of states in 2D. We assume that $\lambda$ is small number.
 The conventional weak coupling BCS analysis is valid when the attraction is confined to energies much smaller than $E_F$, i.e.,  when $\Lambda \ll E_F$.  We consider the opposite situation  when $E_F$ is much smaller than the cutoff energy $\Lambda$.

For two fermions with 2D $k^2$ dispersion in a vacuum (i.e., at $E_F=0$),  an arbitrary small attraction $U$ gives rise  to the formation of a bound state~\cite{LL}. The bound state energy at $T=0$ is $2E_0$, where
\beq
     E_{0} = \Lambda e^{-\frac{2}{\lambda}}
     \label{s_5}
\eeq
The bound state develops at $T_0 = 1.13 E_0$. The 2D  scattering amplitude $a_2 \propto 1/\log{\frac{T_0}{T}}$ diverges at $T=T_0$ and changes sign from negative at $T > T_0$ to positive at $T < T_0$. We consider the system at a non-zero $E_F$, i.e., at a finite density of fermions $n =  m E_F/\pi$.  We
   show that the system behavior  is different at $E_F \gg E_{0}$ and at
     $E_F \ll E_{0}$.\\

\subsection{The onset temperature of the pairing, the pairing gap and the renormalization of the chemical potential}

The onset temperature of the pairing instability, $T_{ins}$ (not necessary a true superconducting transition temperature) is obtained by introducing infinitesimal pairing vertex and dressing it by renormalizations to obtain the pairing susceptibility.  The temperature $T_{ins}$ is the one at which the pairing susceptibility diverges.  To logarithmical accuracy, one needs to keep only ladder series of renormalizations in the particle-particle channel and neglect all  renormalizations coming from  particle-hole channel because the first contain series of $\lambda \log \frac{\Lambda}{T_{ins}}$ while the latter contain series in $\lambda$.  We assume and then verify that in all cases that we consider,  $T_{ins} \ll \Lambda$, hence $\log \frac{\Lambda}{T_{ins}}$ is a large factor. However, as we will see, the ratio of $\frac{T_{ins}}{E_F}$ is small only when $E_F \gg E_0$ and is actually large in the opposite limit $E_F \ll E_0$. Because temperature variation of the chemical potential $\mu (T)$ in the normal state holds in powers of $T_{ins}/E_F$, this variation generally cannot be neglected, i.e., the equation for the pairing vertex at $T = T_{ins}$ has to be combined with the equation for the chemical potential $\mu (T_{ins})$. The latter follows from the condition that the total number of fermions is conserved.~\cite{Mohit89,DeMelo1993}  The two coupled equations are ($\mu = \mu (T_{ins})$)
\bea
1 &=& \frac{\lambda}{2} \int_{0}^\Lambda d {\varepsilon} \frac{\tanh{\frac{\varepsilon-\mu}{2T_{ins}}}}{\varepsilon-\mu} \nonumber\\
&= & \frac{\lambda}{2} \left(\int_{0}^{\mu} d x \frac{\tanh{\frac{x}{2T_{ins}}}}{x} + \int_{0}^{\Lambda} d x \frac{\tanh{\frac{x}{2T_{ins}}}}{x}\right) \nonumber \\
   E_F &=& \int_{0}^\Lambda d {\varepsilon} \frac{1}{e^{(\varepsilon-\mu)/T_{ins}} +1} = T_{ins} \log{(1 + e^{\mu/T_{ins}})} \nonumber\\
  \label{ch_e1}
\eea
At  $E_F \gg  E_{0}$, the solution of these equations yields
\bea
 T_{ins} & = &1.13 (\Lambda E_F)^{1/2} e^{-\frac{1}{\lambda}} \sim \sqrt{E_F E_{0}}  \nonumber \\
 \mu (T_{ins}) & \approx & E_F
 \label{s_3}
 \eea
  In the opposite limit  $E_F \ll E_{0}$ we obtain~\cite{kagan}
  \bea
  T_{ins} & = & \frac{E_0}{\log{\frac{E_0}{E_F}}} \nonumber \\
  \mu (T_{ins}) & \approx & - E_0
  \label{s_4}
  \eea
This behavior is rather similar to that in three-dimensional case.\cite{DeMelo1993}  The behavior of $T_{ins}$ and $\mu$ at intermediate $E_F \sim E_0$  can be easily obtained numerically.

For $E_F=E_0$, $T_{ins} \approx 1.09 E_F$ and $\mu (T_{ins}) \approx -0.09 E_F$. Note that at $E_F \gg E_{0}$, the instability temperature  $T_{ins} \ll E_F$, while at  $E_F \ll E_{0}$, $T_{ins} \gg E_F$ and $\mu (T_{ins})$ is negative.

The prefactor 1.13 in Eq. (\ref{s_3})  is in fact obtained by going beyond logarithmical accuracy in the particle-particle channel.
 To get the correct prefactor one also needs to include fermionic self-energy to order $\lambda$ and the renormalization of $U$  by corrections from the particle-hole channel.~\cite{Gorkov61,extra_1}
These two renormalizations are not essential for our consideration and in the bulk of the text we neglect them. For completeness, however, we obtain the result for $T_{ins}$ with the full prefactor  in the Appendix.

The pairing gap $\Delta$ and the renormalized chemical potential $\mu$ at $T < T_{ins}$ are obtained by solving simultaneously the non-linear gap equation and the equation on
 $\mu (T)$.  The set looks particularly simple at $T=0$ (here $\mu = \mu (T=0),~\Delta = \Delta (T=0)$):
 \bea
1&=&\frac{\lambda}{2} \int_{0}^{\Lambda}d\varepsilon\frac{1}{\sqrt{(\varepsilon -\mu)^2 + \Delta^2}}, \nonumber \\
E_F &=& \frac{1}{2} \int_0^{\infty} d\varepsilon \left(1  - \frac{\varepsilon - \mu}{\sqrt{(\varepsilon-\mu)^2+\Delta^2}} \right).
\label{eq.1band.mu}
\eea
Solving these equations we obtain at $T=0$
 \bea
 &&\mu + \sqrt{\mu^2 + \Delta^2} = 2 E_F \nonumber \\
 &&  \sqrt{\mu^2 + \Delta^2} - \mu = 2 E_0
 \label{s_6aa}
 \eea
 hence
 \bea
 &&\mu = E_F -E_0 \nonumber \\
 &&\Delta = 2 \sqrt{E_F E_0}
 \label{s_6ab}
 \eea
 These results were first obtained in Ref. \cite{Mohit89}.

  When $E_F \gg E_{0}$,  the
   expressions for $\mu$ and $\Delta$ are the same as in BCS theory:
 \bea
 &&\mu \approx E_F, \nonumber \\
 && \Delta = 2 (\Lambda E_F)^{1/2} e^{-\frac{1}{\lambda}} = 1.76 T_{ins}.
\label{s_6a}
 \eea
In the opposite limit $E_F \ll E_0$,
\bea
&&\mu \approx - E_0 \nonumber \\
&&\Delta \sim T_{ins} \left(\frac{E_F}{E_0}\right)^{1/2}  \log{\frac{E_0}{E_F}}.
\label{s_6}
\eea
 Observe that while $\Delta =2 \sqrt{E_F E_0}$ stays the same in both limits, the ratio $\Delta/T_{ins}$ changes: $\Delta \sim T_{ins}$ at $E_F \gg E_0$ and $\Delta \ll T_{ins}$ at $E_F \ll E_0$.  At  $E_F  = 0$, $\Delta$, $T_{ins}$,  and $\Delta/T_{ins}$  all vanish.  The vanishing of $\Delta$
 is easy to understand -- a finite gap would reconstruct fermionic dispersion and open up a hole band with a finite density of carriers proportional to $\Delta$, what is impossible   at $E_F =0$ because the density of fermions is zero.  A negative $\mu$  implies that  the Fermi momentum $k_F$ (defined as position of the minimum of $E_k = \sqrt{(\varepsilon_k - \mu)^2 + \Delta^2}$) is zero.  In fact, the Fermi momentum shifts downwards already in the normal state at a finite $T$ because    $\mu (T) < E_F$.  It becomes zero at $T = T_{ins}$ at $E_0/E_F \approx 0.882$.  The downward renormalization of $k_F$  has been recently obtained  in the study of superconductor-insulator transition.~\cite{Mohit2015}

\begin{figure}[t!]
\renewcommand{\baselinestretch}{1.0}
\includegraphics[angle=0,width=.8\linewidth]{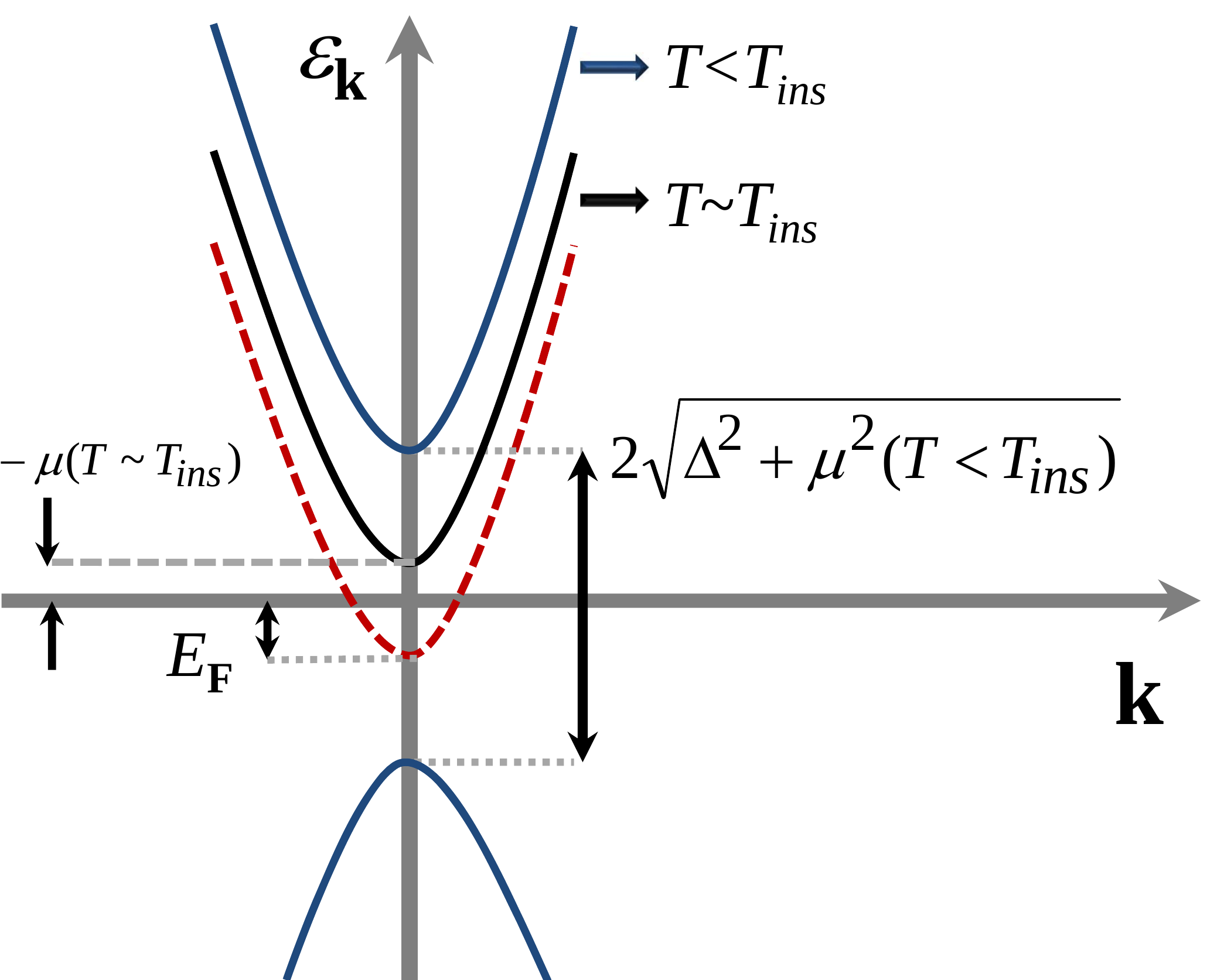}
\caption{The dispersion in the one-band model. Red dashed line -- the bare dispersion (the one which the system would have at $T=0$ in the absence of the pairing). Black line -- the dispersion right above $T_{ins}$, blue line -- the dispersion below $T_{ins}$.  The plot is for the case when the chemical potential $\mu$ is already negative at $T= T_{ins}$. Observe that the minimal gap is $\sqrt{\Delta^2 + \mu^2}$ and the minimum of the dispersion is at $k=0$ rather than at $k_F$.}
\label{fig:bands_1}
\end{figure}

In Fig. \ref{fig:bands_1} we plot  the actual dispersion below and above $T_{ins}$ along with the bare fermionic dispersion (the one which the system would have at $T=0$ in the absence of the pairing).

\subsection{Superconducting $T_c$}
\label{sec:sc_one_band}

 The temperature $T_{ins}$ appears  in the ladder approximation as the transition temperature, but is actually only the crossover temperature as pair formation by itself does not break the gauge symmetry. To obtain the actual $T_c$, at which the gauge symmetry  is broken (i.e., the phases of bound pairs order),  one needs to treat the phase $\phi (r)$ as fluctuating variable and compute the energy cost of phase variation $\delta E = (1/2)\rho_s (T) \int dr \left|\nabla \phi\right|^2$ (see Ref. [\onlinecite{LV}] for a generic description of fluctuations in superconductors).
  The prefactor $\rho_s (T)$ is the superfluid stiffness.   In 2D,  superconducting $T_c$ (the temperature of Berezinsky-Kosterlitz-Thouless transition) is of order $\rho_s (T=0)$ (Ref. \cite{Pokrovsky1979,beasley}), provided that $\rho_s (T=0)$ is smaller than $T_{ins}$.  If $\rho_s (T=0) \gg T_{ins}$,  phase fluctuations cost too much energy, and the phases of bound pairs order almost immediately after the pairs develop. In this last case  $T_c = T_{ins}$ minus a small correction.
\begin{figure}[t!]
\renewcommand{\baselinestretch}{.8}
\includegraphics[angle=0,width=.8\linewidth]{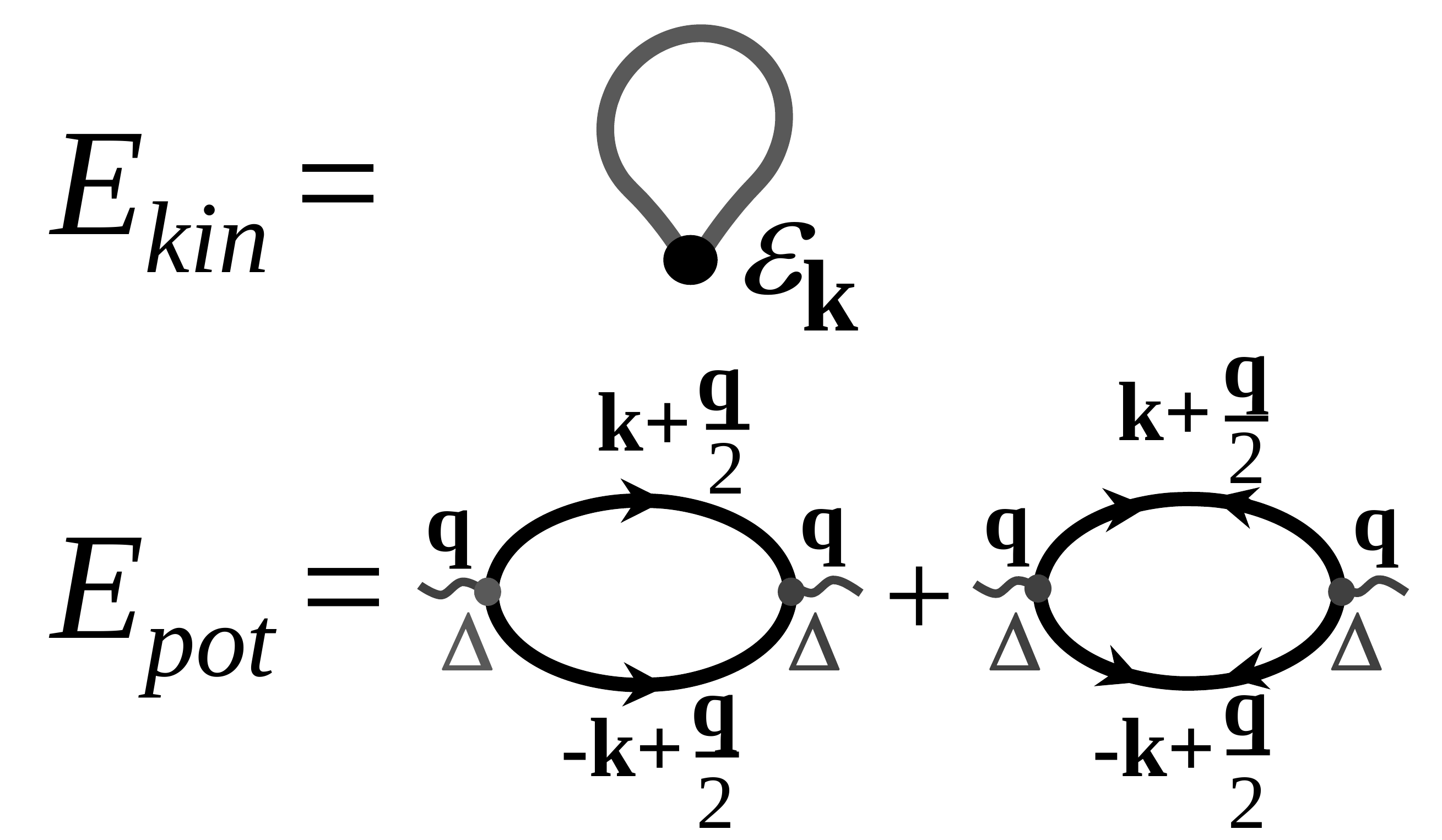}
\caption{Diagrammatic representation of the kinetic and potential energy of a one-band superconductor. The sum $E_{kin} + E_{pot} (q=0)$ gives the condensation energy, and the prefactor for the $q^2$ term in $E_{pot} (q)$ determines the superfluid stiffness.}
\label{fig3}
\end{figure}

 Within our model with local interaction $U$, $\delta E$ is the $O(q^2)$  term in the ground state energy of an effective model described by the effective fermionic Hamiltonian with  the anomalous term
 \bea
 {\cal H}_{anom} & = &  \int d^2 r \Delta (r) c_{\uparrow}^\dagger (r) c_{\downarrow}^\dagger (r) + h.c  \nonumber \\
 &=& \sum_{k,q} \Delta (q) c^\dagger_{\uparrow} (k+q/2) c_{\downarrow}^\dagger (-k+q/2) + h.c \nonumber \\
 \label{ch_e7}
  \eea
  with $\Delta (r) = \Delta e^{i\phi (r)} \approx \Delta e^{i (\nabla \phi)r}$ whose Fourier componet $\Delta (q) =\Delta \delta (q-\nabla \phi)$.

    The  ground state energy is the sum of the kinetic and the potential energy. The kinetic energy depends on $|\Delta (r)|^2 = \Delta^2$ and is not sensitive to phase fluctuations (i.e., it does not have $(\nabla \phi)^2$ term) and is simply given by the convolution of the quasiparticle dispersion  with a single fermionic Green's function (see Fig.\ref{fig3}).
 At $T=0$
 \bea
 E_{kin} & = & 2N_0 \int \frac{d \varepsilon_k d \omega}{2\pi}
  \varepsilon_k G_s(k, \omega) = \nonumber \\
  && -2N_0 \int \frac{d \varepsilon_k d \omega}{2\pi}
  \varepsilon_k \frac{i \omega + (\varepsilon_k -\mu)}{\omega^2 + (\varepsilon_k -\mu)^2 + \Delta^2}
 \label{ch_e6}
 \eea
 where  $G_s$ is the normal of the superconducting Green's function.

 The potential energy, on the other hand, does depend on $q$.  Within the model of Eq. (\ref{ch_e7}) it is given by the sum of the convolutions of two normal and two anomalous Green's functions with $\Delta$ in the vertices\cite{tobij2000,LV,scal} (see Fig. 3).   In the analytic form
 \begin{widetext}
\bea
E_{pot} (q)  =  -\Delta^2 \int \frac{d^2 k d \omega}{(2\pi)^3} \left[G_s (k+q/2,\omega) G_s (-k+q/2, -\omega) +  F_s (k+q/2,\omega) F_s (-k+q/2, -\omega)\right],
 \label{ch_e8}
 \eea
 \end{widetext}
 where $F_s$ is the anomalous Green's function.
 Integrating over frequency in Eq.(\ref{ch_e8}) we obtain
 at $T=0$
  \bea
 \lefteqn{E_{pot} (q) = - \frac{\Delta^2}{|U|} +  \frac{\Delta^2}{4} \int d^2 k \frac{(\varepsilon_{k+q/2} - \varepsilon_{k-q/2})^2}{\left((\varepsilon_k-\mu)^2 + \Delta^2\right)^{3/2}} +...} \nonumber \\
  &=& - \frac{\Delta^2}{|U|} + q^2 \frac{\Delta^2}{8\pi} \int d\varepsilon_k \frac{\varepsilon_k}{\left((\varepsilon_k-\mu)^2 + \Delta^2\right)^{3/2}} + ...
   \label{ch_e9}
 \eea
where dots stand for the terms of higher orders in $q^2$.
The difference between $E_{pot} (q=0) + E_{kin}$  in a superconductor and the kinetic energy in the normal state gives the
condensation energy  $E_{cond}$.
To obtain $E_{cond}$  we evaluate the frequency integrals in (\ref{ch_e6}) and (\ref{ch_e9}) and write the condensation energy as
\bea
\lefteqn{E_{cond} = -N_0 \times}\nonumber\\
&&\left[E^2_F  + \frac{\Delta^2}{2} \int^\Lambda_{-\mu} \frac{dx}{\sqrt{x^2 + \Delta^2}} \left(1-\frac{2(x+\mu)}{x +\sqrt{x^2 + \Delta^2}}\right)\right] \nonumber\\
\eea
The integral over $x$ is ultra-violet convergent and one can safely replace the upper limit by infinity. We then  obtain
\beq
E_{cond} = - N_0 \left[E^2_F + \frac{\Delta^2}{4} - \frac{\mu}{2} \left(\mu + \sqrt{\mu^2 + \Delta^2}\right)\right]
\label{yyy}
\eeq
Using $\mu = E_F - E_0$ and $\Delta^2 = 4 E_F E_0$ (see Eq. (\ref{s_6ab})) we immediately obtain $\mu + \sqrt{\mu^2 + \Delta^2} = 2 E_F$ and
 $ \mu\left(\mu + \sqrt{\mu^2 + \Delta^2}\right)/2 = E^2_F - \Delta^2/4$.  Substituting into (\ref{yyy}) we obtain
 \beq
 E_{cond} = -N_0 \frac{\Delta^2}{2} = - N_0 E_0 E_F
 \eeq
  no matter what the ratio $E_F/E_0$ is.

 The prefactor for the $q^2$ term in Eq.(\ref{ch_e9})  determines $\rho_s (T=0)$:
\beq
\rho_s (T=0) = N_0 \frac{\Delta^2}{8} \int d\varepsilon_k \frac{\left(\frac{d\varepsilon_k}{dk}\right)^2}{\left((\varepsilon_k-\mu)^2 + \Delta^2 \right)^{3/2}}
\label{ch_e10}
\eeq
Using
\beq
\frac{\Delta^2}{\left((\varepsilon_k-\mu)^2 + \Delta^2\right)^{3/2}} = -\frac{d}{d \varepsilon_k} \left(1 - \frac{\varepsilon_k-\mu}{\left((\varepsilon_k-\mu)^2 + \Delta^2\right)^{1/2}}\right)
\eeq
 and integrating by parts Eq.(\ref{ch_e10}) we obtain
\bea
\lefteqn{\rho_s (T=0) =  \frac{N_0}{8} \int d\varepsilon_k \times}\nonumber \\
&& \left(1 - \frac{\varepsilon_k-\mu}{\left((\varepsilon_k-\mu)^2 + \Delta^2\right)^{1/2}}\right)  \left[\frac{d}{d \varepsilon_k} \left(\frac{d \varepsilon_k}{d k}\right)^2\right]
\label{ch_e10a}
\eea
 The term in square brackets is simply a constant ($=2/m$), and the remaining integral gives exactly the total energy density equal to $2 E_F$. As a result,
 \beq
 \rho_s = \frac{E_F}{4\pi}
 \eeq
Note that this result is exact for the Galilean invariant case when the dispersion is exactly $\frac{k^2}{2m}$, but also holds, up to corrections of order $(E^2_0 + E^2_F)/\Lambda^2$, for arbitrary lattice dispersion.~\cite{mohit}

At $E_F \gg E_0$, $\rho_s$ is parametically larger than $T_{ins} \sim (E_F E_0)^{1/2}$. As the consequence, phase fluctuations are costly and $T_c \approx T_{ins}$, i.e., fermionic pairs condense almost immediately after they develop.
 In the opposite limit $E_F \ll E_0$, $\rho_s (T=0)  \ll T_{ins}$, and hence $T_c \sim \rho_s (T=0)  \ll T_{ins}$.
 Using the criterium~\cite{Pokrovsky1979} $T_c = (\pi/2) \rho_{s} (T)$
 and approximating $\rho_{s} (T)$ by $\rho_{s} (T=0)$,  we obtain an estimate $T_c =E_F/8$. (A more accurate analysis~\cite{more_accurate} yields $T_c \sim E_F/\log{(\log{E_0/E_F})}$.)

\begin{figure}[t!]
\renewcommand{\baselinestretch}{1.0}
\includegraphics[angle=0,width=1.1\linewidth]{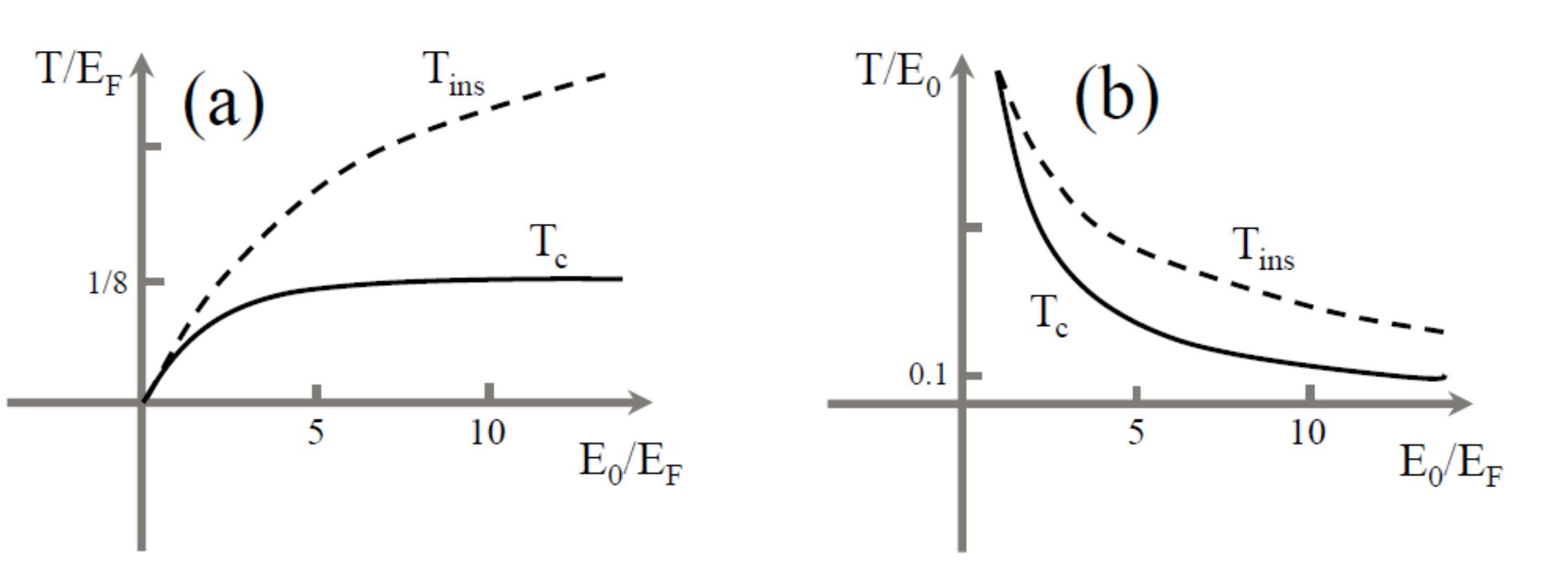}
\caption{The  onset temperature $T_{ins}$ for the bound state formation and the superconducting transition temperature $T_c$ in the one-band model
as  functions of $E_0/E_F$. The temperatures are  normalized to $E_F$ (a) and to $E_0$ (b).  Observe that $T_{ins}$ scales as $E_0/\log{E_0/E_F}$ at large $E_0/E_F$. This $T_{ins}$ increases when plotted in units of $E_F$ and decreases when plotted in units of $E_0$.}
\label{fig4n}
\end{figure}

The superconducting transition temperature approaches zero as $O(E_F)$ when $E_F \to 0$, while $T_{ins} \sim E_0/\log\frac{E_0}{E_F}$ drops only logarithmically.
The ratio  $T_c/T_{ins}$ scales as $\frac{E_F}{E_0} \log{\frac{E_0}{E_F}}$ and obviously vanishes when $E_F =0$.  In the temperature region between $T_{ins}$ and $T_c$ the bound pairs develop but remain incoherent.  In Fig.\ref{fig4n} we plot $T_{ins}$ and $T_c$ as functions of $E_0/E_F$.

 The splitting between $T_{ins}$ and $T_c$ once $E_F$ gets smaller than $T_{ins}$  (BCS-BEC crossover) and the corresponding preformed pairs behavior at $T_{ins} > T > T_c$  has been originally studied in 3D systems.~\cite{DeMelo1993,DeMelo2006,Ohashi2003} The physics in 2D is  similar, but there is one important difference --
    the distance between fermions in a bound pair (the coherence length $\xi_0$) scales as $\xi_0 \sim \sqrt{\frac{|\mu|}{m}}/\Delta \sim 1/k_F$, while
     the interatomic distance $a_0 \sim 1/(m \Lambda)^{1/2} \sim 1/q_{max}$, where $q_{max} = (2m \Lambda)^{1/2}$.
      The ratio $\xi_0/a_0 \sim (\Lambda/E_F)^{1/2}$ is large, hence
       fermions in a bound pair are on average located much farther away from each other than interatomic spacing. Hence, the pairs cannot be viewed as "molecules" in the real space. In this respect our result differs from the analysis in Ref.~\cite{Mohit89}, where it was argued that at $E_F \ll E_0$, $\xi_0$ becomes much smaller than the interatomic spacing.

Our results for the one-band model differ from Refs. \cite{Chen2015,phillips2015} where $T_c$ was found to remain finite at $E_F=0$.  The authors of \cite{Chen2015,phillips2015}
 solved BCS-like equations, hence their $T_c$ is in fact the onset temperature for the pairing, $T_{ins}$. Still, we found that even this temperature vanishes at $E_F =0$, once one includes into consideration temperature variation of the chemical potential.

 \subsection{The density of states at $T=0$}

 In a conventional BCS superconductor with $E_F \gg T_c$, $\mu (T=0)$ is positive, and  the density of states (DOS) at $T=0$ is, for electron dispersion
 \begin{widetext}
 \bea
 &&N(\omega) = -\frac{1}{\pi} \mbox{Im}  \int \frac{d^2 k}{4\pi^2} G_s(k, \omega) =  Im \int \frac{d^2 k}{4\pi^2} \left({u^2_k} \delta(\omega- E_k) + {v^2_k} \delta(\omega+ E_k)\right) \nonumber \\
 &&= \frac{N_0}{2}\left(\frac{2 \omega}{\sqrt{\omega^2-\Delta^2}} \theta (\omega -\Delta) - \frac{\omega - \sqrt{\omega^2-\Delta^2}}{\sqrt{\omega^2-\Delta^2}} \theta\left(\omega - \sqrt{\mu^2 +\Delta^2}\right)\right)  \nonumber \\
 &&+ \frac{N_0}{2}\left(\frac{2 |\omega|}{\sqrt{\omega^2-\Delta^2}} \theta (-\omega -\Delta) - \frac{|\omega| + \sqrt{\omega^2-\Delta^2}}{\sqrt{\omega^2-\Delta^2}} \theta\left(-\omega - \sqrt{\mu^2 +\Delta^2}\right) \right),
 \label{nn_a}
 \eea
 \end{widetext}
where $E_k = \sqrt{(\varepsilon_k -\mu)^2 + \Delta^2}$, $N_0 = m/(2\pi)$, $\theta (x) = 1$ for $x >0$, $\mu = \mu (T=0)$,
  and
\bea
&&u^2_k = \frac{1}{2} \left(1 + \frac{\varepsilon_k -\mu}{\sqrt{(\varepsilon_k -\mu)^2 + \Delta^2}}\right), \nonumber \\
&& v^2_k = \frac{1}{2}
 \left(1 - \frac{\varepsilon_k -\mu}{\sqrt{(\varepsilon_k -\mu)^2 + \Delta^2}}\right),
 \eea
 This $N (\omega)$ vanishes at $|\omega| < \Delta$, has a square-root singularity $1/\sqrt{|\omega -\Delta|}$ above the gap,  and
  drops by a finite amount at $|\omega| =  \sqrt{\mu^2 +\Delta^2} +0$, when $|\omega|$ crosses the edge of the band.  The DOS is
  nearly symmetric between positive and negative $\omega$, at least for $\Delta < \omega \ll E_F$.

\begin{figure}[t!]
\renewcommand{\baselinestretch}{1.0}
\includegraphics[angle=0,width=1.0\linewidth]{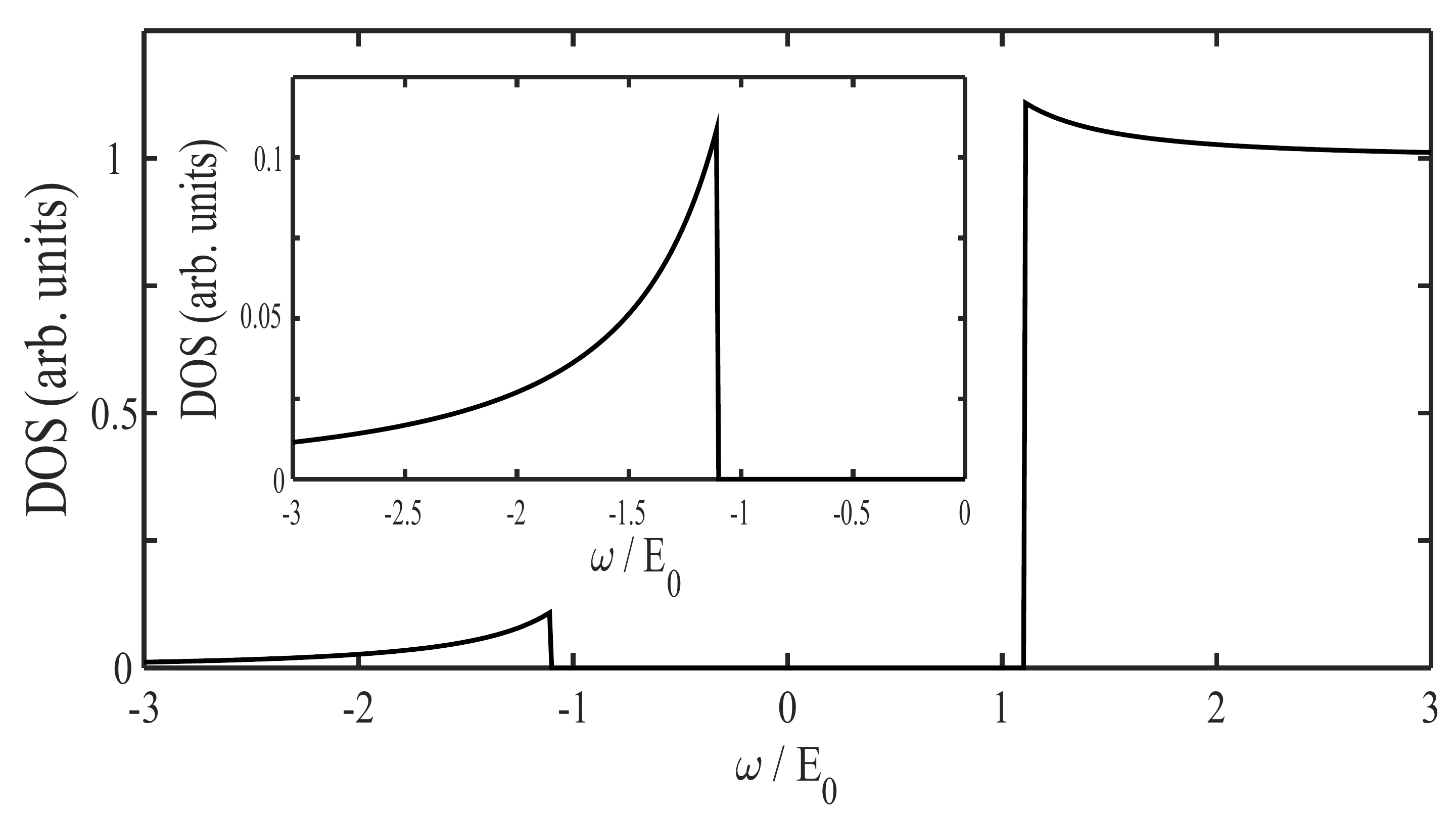}
\caption{The DOS in the single-band model at $T=0$ for $E_F\ll E_0$.   We set
 $E_F=0.1 E_0$, in which case $\mu = -9 E_F$ and  $\Delta=2 \sqrt{E_0 E_F} \approx 6.3 E_F$.
   We
    added the fermionic damping $\gamma=0.001 E_F$.  In the clean limit, the density of states vanishes at $|\omega| < \sqrt{\mu^2+\Delta^2}$ and
      jumps to a finite value at  $|\omega| = \sqrt{\mu^2+\Delta^2} +0$ (see Eq. (\protect\ref{nn_b})).  Due to the difference between coherence factors,  the DOS is strongly particle-hole anisotropic and has much larger value at positive
       frequencies, unobservable in photoemission experiments.
        At large $|\omega|$, the DOS tends to a finite value at positive $\omega$ and vanishes as $1/\omega^2$ at negative $\omega$.
    To make the power-law suppression of the DOS at negative $\omega$
    more visible,  we plot the negative frequency region separately in the inset.}
\label{fig_dossingleband_new}
\end{figure}

   In our case, this behavior holds for the case $E_F \gg E_0$, but not for $E_F \ll E_0$.  In the latter case, $\mu (T=0)$ is negative ($\mu (T=0) \approx - E_0$),
   and the DOS is given by
   \bea
 &&N(\omega) = -\frac{1}{\pi} \mbox{Im}  \int \frac{d^2 k}{4\pi^2} G_s(k, \omega) \nonumber \\
 && =  Im \int \frac{d^2 k}{4\pi^2} \left({u^2_k} \delta(\omega- E_k) + {v^2_k} \delta(\omega+ E_k)\right) \nonumber \\
 &&= \frac{N_0}{2}\left(\frac{\omega + \sqrt{\omega^2-\Delta^2}}{\sqrt{\omega^2-\Delta^2}}\right) \theta\left(\omega - \sqrt{\mu^2 +\Delta^2}\right) \label{nn_b} \\
 &&+ \frac{N_0}{2}\left( \frac{|\omega| - \sqrt{\omega^2-\Delta^2}}{\sqrt{\omega^2-\Delta^2}}\right) \theta\left(-\omega - \sqrt{\mu^2 +\Delta^2}\right), \nonumber
 \eea
 where $\mu = \mu (T=0)$.  This $N(\omega)$ vanishes at $|\omega| < \sqrt{\mu^2 +\Delta^2}$ and jumps to a finite value at $|\omega| = \sqrt{\mu^2 +\Delta^2} +0$.
  Because    $\mu \approx - E_0$, is much larger than
 $\Delta = 2(E_F E_0)^{1/2}$, the coherence factors $u^2_k$
  and $v^2_k$  are quite different: $u^2_k \approx 1$ for all momenta, while $v^2_k \approx \frac{\Delta^2}{4 (\varepsilon_k + |\mu|)}$ is small.  As the result,
    $N(\omega)$  in (\ref{nn_b})
    is highly anisotropic between positive and negative frequencies -- it is approaches $N_0$ at large positive frequencies
     and scales as  $N_0 \Delta^2/(4 \omega^2)$ at large negative frequencies.
 We plot the DOS  at zero temperature for $E_F \ll E_0$ in Fig.  \ref{fig_dossingleband_new}.
     Because only negative frequencies are probed in photoemission experiments, the features associated with the bound state development below $T_{ins}$ are weak and disappear at $E_F \to 0$. This last feature has been also found in the recent study of superconductor-insulator transition.~\cite{Mohit2015}

Note in passing that within our approximate treatment, based on the effective quadratic Hamiltonian,  the DOS vanishes below $|\omega| =  (\mu^2 + \Delta^2)^{1/2}$ already
   at $T < T_{ins}$.  A more accurate treatment would require one to compute the imaginary part of the fermionic self-energy at a finite temperature
     and analyze the feedback on this self-energy from the development of the bound pairs. On general grounds, the density of states at $|\omega| <  (\mu^2 + \Delta^2)^{1/2}$
     should remain  finite at temperatures between $T_{ins}$ and $T_c$, as no symmetry is broken in this $T$ range.  Below $T_c$, however, the true gap develops at these frequencies and the DOS should be as in Fig. \ref{fig_dossingleband_new}.

  \section{Two-band model with one hole and one electron band}
\label{sec_3}

We now extend the  analysis to two-band models. We  consider two models -- one with a hole and an electron band, and  one with
 two hole/two electron bands
 In both cases we assume, to make presentation compact, that the dominant pairing interaction $U(q, \Omega)$ is the pair hopping between
   fermions on the two bands.  The repulsive interaction of this type gives rise to $s^{+-}$ pairing with the phase shift by $\pi$ between $\Delta$'s on the two bands.

In this section we consider the model with one
 band with hole-like dispersion $\varepsilon^h_k = E_{F,h} - \frac{k^2}{2m_h}$ and another with electron-like dispersion $\varepsilon^e_k = \frac{k^2}{2m_e} - E_{F,e}$.
 This model is relevant to FeSCs, at least, at a qualitative level. The Fermi energies $E_{F,h}$  and $E_{F,e}$ and the masses $m_h$ and $m_e$ are generally not equivalent. We keep
   $E_{F,h}$ and $E_{F,e}$ different, but set $m_h = m_e =m$ to simplify the formulas.

BCS analysis of the pairing in  multi-band  models with two electron bands and two or three hole bands, as in FeSCs,
   has been presented in series of recent publications~\cite{Chen2015,phillips2015}.
 In particular, Ref. \cite{Chen2015} considered  the case of two hole bands, only one of which crosses the Fermi level.
   A FS-constrained superconductivity in this last case emerges due to interaction between  the hole band with $E_{F,h} >0$ and the electron band(s). Ref. \cite{Chen2015} has demonstrated that the presence of the additional hole band increases $T_c$, despite that this band is full located below the Fermi level.

We analyze  different physics -- the crossover in the system behavior once the largest $E_F$ becomes smaller than the two-particle bound state energy $E_0$.
 This physics has not been analyzed before, to the best of our knowledge.
  We restrict to one hole and one electron band  because the inclusion of additional bands affects the details of the analysis but does not qualitatively affect BCS-BEC crossover.
  Like in the previous Section, we set the upper energy cutoff at $\Lambda \gg E_{F,i}$ ($i =h,e$), approximate $U(q,\Omega)$ by a constant below the cutoff,
   and set the dimensionless coupling coupling $\lambda = m U/(2\pi)$ to be small.  We set $U >0$, in which case superconducting order parameter has $s^{+-}$ symmetry.
 As our goal is to analyze BCS-BEC crossover,  we consider the particular case when $E_{F, e} =0$, see Fig.\ref{fig1}(b). The extension of the analysis to small but finite $E_{F, e}$ (positive or negative) is straightforward and does not bring qualitatively new physics.

 The analysis of the bound state energy for two particles at $E_F \equiv 0$ does not differ from that in previous Section, and the result is that  the scattering amplitude diverges at $T_0 = 1.13 E_0$, where, like before,  $E_0 = \Lambda e^{-2/\lambda}$.  The bound state energy at $T=0$ is $2 E_0$.

The onset temperature for the pairing at a finite $E_F$  is obtained by solving simultaneously the linearized  equations for $\Delta_e$ and $\Delta_h$ and the equations for the chemical potentials $\mu_e (T)$ and $\mu_h (T)$, subject to $\mu_e (T) + \mu_h (T) = E_F$. The equation for the chemical potential
follows for the conservation of the total number of fermions.  The set of equations is ($\mu_e = \mu_e (T_{ins}), \mu_h = \mu_h (T_{ins})$):
 \bea
\Delta_e &=& - \frac{\lambda}{2} \Delta_h \int_{-\mu_h}^{\Lambda}  \frac{dx}{x} \tanh{\frac{x}{2T_{ins}}} \nonumber \\
\Delta_h &=& -  \frac{\lambda}{2} \Delta_e \int_{-\mu_e}^{\Lambda}  \frac{dx}{x} \tanh{\frac{x}{2T_{ins}}} \nonumber \\
\mu_e &=& T \log{\frac{1 + e^{-\mu_h/T}}{1 + e^{\mu_e/T}}} \nonumber \\
\mu_e &+& \mu_h = E_F.
\label{ch_11}
\eea
The first two equations reduce to
\beq
\frac{4}{\lambda^2} =  \int_{-\mu_e}^{\Lambda}  \frac{dx}{x} \tanh{\frac{x}{2T_{ins}}} \times \int_{-\mu_h}^{\Lambda}  \frac{dx}{x} \tanh{\frac{x}{2T_{ins}}}
\label{ch_11_0}
\eeq

 Below $T_{ins}$,  $\Delta_e$ and $\Delta_h$ become non-zero and one has to consider non-linear gap equations and modify the equation for the chemical potential.
 At $T=0$ the set of equations becomes [$\mu_h = \mu_h (T=0), \mu_e = \mu_e (T=0)$]:
\bea
&&\Delta_h = - \frac{\lambda}{2}  \Delta_e \log{\frac{2\Lambda}{\sqrt{\mu^2_e+ \Delta^2_e} - \mu_e}}\nonumber \\
&&\Delta_e = - \frac{\lambda}{2}  \Delta_h \log{\frac{2\Lambda}{\sqrt{\mu^2_h+ \Delta^2_h} - \mu_h}} \nonumber \\
&& \sqrt{\mu^2_h+ \Delta^2_h} + \mu_h -2E_F=  \mu_e + \sqrt{\mu^2_e + \Delta^2_e} \nonumber \\
&& \mu_h + \mu_e = E_F
\label{ch_11_1}
\eea

 \subsection{The case $E_F  \gg E_0$}
\label{sec_3a}

We assume and then verify that in this situation  $T_{ins} \ll E_F$ and for all $T \leq T_{ins}$, $\mu_h \approx E_F$, while $\mu_e \sim T_{ins}$
   Under these assumptions, the equations on the chemical potentials in (\ref{ch_11}) yield
\bea
\mu_e (T_{ins}) & = &  T_{ins} \log{\frac{\sqrt{5}-1}{2}} = -0.48 T_{ins}, \nonumber\\
\mu_h (T_{ins}) & = & E_F - \mu_e (T_{ins})
\label{fri_4}
\eea
Substituting these values of the chemical potentials into the first two equations in (\ref{ch_11}) we obtain after simple algebra
\bea
\Delta_e &=& - \frac{\lambda}{2}  \Delta_h \left[2\log{\frac{1.13 \Lambda}{T_{ins}}} - \frac{2}{\lambda} + \log{\frac{E_F}{E_0}}\right] \nonumber \\
\Delta_h &=& -  \frac{\lambda}{2}  \Delta_e \log{\frac{1.13 D \Lambda}{T_{ins}}}
\label{ch_11_2}
\eea
 where $D = 0.79$  ($\log D = -\int_0^{|\mu_e|/2T} dx \frac{\tanh{x}}{x}$).
Combining the two equations and introducing $Z = \log{\frac{1.13 D \Lambda}{T_{ins}}}$, we obtain at small $\lambda$,
\beq
Z =  \frac{2}{\lambda} - \frac{1}{3} \log{\frac{E_F}{D^2 E_0}} + O(\lambda).
\eeq
Hence
\beq
T_{ins} = 1.13 D^{1/3} E^{1/3}_F E^{2/3}_0 = 1.04 E^{1/3}_F E^{2/3}_0
\label{ch_12}
\eeq
This expression is valid when $\lambda \log{E_F/E_0} \ll 1$.  At even larger $E_F \leq \Lambda$, when $\lambda \log{E_F/E_0} = O(1)$,  $T_{ins}$ is given by
\beq
T_{ins} \sim \Lambda \left(\frac{E_F}{\Lambda}\right)^{1/4} e^{-\frac{\sqrt{2}}{\lambda}}
\label{ch_14}
\eeq

The ratio of the gaps on electron and hole bands at $T = T_{ins} -0$ is
\beq
\frac{\Delta_e}{\Delta_h} \approx -\left(1 + \frac{\lambda}{6} \log{\frac{E_F}{E_0}}\right)
\label{ch_15}
\eeq
We see that the gap on the electron band, which touches the Fermi level,
is {\it larger} than the gap on the hole band, which crosses Fermi level.  This result indeed holds even when $E_F$ becomes negative, i.e., the electron band is above the Fermi level.  The ratio of $\Delta_e/\Delta_h$ increases when $E_F$ gets larger  and approaches $\sqrt{2}$ when $E_F$ becomes of order $\Lambda$.

At $T=0$,  solution of the set (\ref{ch_11_1}) at $E_F \gg E_0$ but $\lambda \log{E_F/E_0} \ll 1$ shows that the
 ratio of $\Delta_e$ and $\Delta_h$ remains the same as in Eq.  (\ref{ch_15}), i.e., up to subleading terms
 $\Delta_e (T=0) = -\Delta_h (T=0) = \Delta$.
 Solving for the chemical potentials we then find
 \beq
 \mu_e (T=0) = -\frac{\Delta}{2\sqrt{2}}, ~~\mu_h (T=0) = E_F - \mu_e (T=0)
 \eeq
 Substituting this into the first two equations in Eq. (\ref{ch_11_1}) and solving for $\Delta$ we obtain
 \beq
 \Delta = 2^{5/6} E^{1/3}_F E^{2/3}_0 = 1.78 E^{1/3}_F E^{2/3}_0
 \eeq

The minimum energy on the hole band is $E_h = \Delta$, at $k \approx k_F$.  The  minimum energy on the electron band is at $k=0$, and
$E_e = \sqrt{\mu^2_e (T=0) + \Delta^2} = 3 \Delta/2\sqrt{2} = 1.06 \Delta$. For the ratio of the minimal energy to $T_{ins}$ we then have, up to corrections of order $\lambda \log {E_F/E_0}$,
\beq
\frac{E_e}{T_{ins}} = 1.71, ~~ \frac{E_h}{T_{ins}} = 1.81
\eeq
Note that both ratios are rather close to BCS values, although our consideration includes the renormalization of the chemical potential, neglected in BCS theory.

\subsection{The case $E_F = E_0$}
\label{sec_3c}

To establish the bridge to the case of small $E_F/E_0$,  consider the intermediate case
  when $E_F$ is comparable to $E_0$.  To be specific, we just set $E_F=E_0$, although the analysis can be easily extended to arbitrary $E_F/E_0 \sim O(1)$.
   Because $E_F$ is now the only relevant low-energy scale,  we express
  $T_{ins} = a E_F, \mu_e (T_{ins}) = b E_F, \mu_h (T_{ins})= E_F (1-b)$.
  Substituting this into Eq.(\ref{ch_11}) and using the fact that
  \beq
  \frac{2}{\lambda} \int_0^\Lambda \frac{\tanh{\frac{x}{2T_{ins}}}}{x} = \frac{2}{\lambda} \log{1.13 \Lambda/T_{ins}} = 1 - \frac{2}{\lambda} \log{\frac{a}{1.13}},
  \eeq
 we obtain, to leading order in $\lambda$, the set of two equations on the prefactors $a$ and $b$:
\bea
&&b = a \log{\frac{1 + e^{\frac{b-1}{a}}}{1 + e^{\frac{b}{a}}}} \nonumber \\
&&2 \log{\frac{a}{1.13}} = \int_0^{\frac{b}{2a}} dy \frac{\tanh{y}}{y} + \int_0^{\frac{1-b}{2a}} dy \frac{\tanh{y}}{y}
\label{ch_16aa}
\eea
Solving the set we obtain $T_{ins} = 1.351 E_F, \mu_e  (T_{ins}) = -0.349 E_F$ and $\mu_h (T_{ins}) = 1.349 E_F$.   As expected,
 the chemical potential $\mu_e$ becomes negative at a finite temperature.

At $T=0$ the  renormalized chemical potentials $\mu_h (T=0)$ and $\mu_e (T=0)$ and the gaps $\Delta_h$ and $\Delta_e$ are also of order $E_F$.
  We express   $\Delta_h = c_h E_F, \Delta_e = c_e E_F$, $\mu_e (T=0) = {\bar b} E_F, \mu_h (T=0) = E_F (1-{\bar b})$.
  Substituting into Eq. (\ref{ch_11_1}) we obtain to leading order $\lambda$, $c_h = -c_e =c$, i.e., $\Delta_e \approx -\Delta_h$.
  [For non-equal masses $m_h$ and $m_e$, $\Delta_h = \Delta \left(m_e/m_h\right)^{1/4}, ~ \Delta_e = - \Delta \left(m_h/m_e\right)^{1/4}$].
  The prefactors $c$ and ${\bar b}$ are the solutions of
 \bea
 &&\left(\sqrt{{\bar b}^2+c^2} -{\bar b}\right)*\left(\sqrt{(1-{\bar b})^2 +c^2} - (1-{\bar b})\right) =4 \nonumber \\
 && \sqrt{(1-{\bar b})^2 +c^2} - \sqrt{{\bar b}^2+c^2} = 1+2{\bar b}
 \eea
 Solving this set we find ${\bar b} = - 0.34$ and $c = 2.43$, i.e., $\mu_e (T=0) = - 0.34 E_F, \mu_h (T=0) = 1.34 E_F$, and $\Delta_h \approx -\Delta_e = 2.43 E_F$. We see that $\mu_e$ and $\mu_h$ change little between $T= T_{ins}$ and $T=0$.

 Because $\mu_e$ is negative, the minimal excitation energy for the electron band is (for $m_h=m_e$) $E_e = \sqrt{\mu^2_e + \Delta^2} \approx 2.45 E_F$.
 For the hole band, $\mu_h$ positive and the minimal energy is $E_h = \Delta$.   We emphasize that the minimal energy $E_e$ is  larger than $E_h$, despite
  that the gaps $\Delta_e$ and $\Delta_h$ have equal magnitudes.
 The ratios of the minimal energy and $T_{ins}$ are
 \beq
 \frac{E_e}{T_{ins}} = 1.82, ~~ \frac{E_h}{T_{ins}} = 1.80
 \eeq
 Both are a bit larger than the  BCS value of 1.76.

 \subsection{The case $E_F \ll E_0$}
 \label{sec_3b}

We assume and then verify that in this limit  the onset temperature for the pairing $T_{ins}$ and the gaps $\Delta_h$ and $\Delta_e$  become progressively larger than $E_F$, while $\mu_h$ and $\mu_e$ remain of order $E_F$.
Assuming that $T_{ins} \gg E_F$ and solving (\ref{ch_11}) for $T_{ins}$,   we then immediately obtain
 \bea
 T_{ins} &=& 1.13 \Lambda e^{-\frac{2}{\lambda}} + \frac{E_F}{4} + O(\lambda) \approx 1.13 E_0\left(1+  0.22\frac{E_F}{E_0}\right) \nonumber \\
 && = T_0 \left(1+  0.22\frac{E_F}{E_0}\right)
 \label{ch_16}
 \eea
 Note that this  differs from  BCS formula because  the exponent contains $2/\lambda$ rather than $1/\lambda$.  The reason is that only fermions with energies above $E_F$ contribute to the logarithm.
  Solving the last two equations from Eq. (\ref{ch_11}), we obtain for the chemical potentials at $T = T_{ins}$
  \beq
  \mu_e (T_{ins}) \approx - \frac{E_F}{2},~~ \mu_h (T_{ins}) \approx \frac{3E_F}{2}
  \eeq

 Solving next for the gaps and the chemical potentials at $T=0$  we obtain (for $m_h = m_e$) that the $\mu_e$ and $\mu_h$  move only little below $T_{ins}$,
   while $\Delta_h \approx - \Delta_e = \Delta$ is related to $T_{ins}$ by the same formula as in BCS theory, i.e.,
  \bea
&& \Delta = 1.76 T_{ins} \gg E_F \nonumber \\
&& \mu_h \approx \frac{3E_F}{2},  ~~ \mu_e \approx -\frac{E_F}{2}
\label{ch_16_1}
\eea
The results for $T_{ins}, \Delta$, and the chemical potential are all consistent with what we assumed a-priori, hence the computational procedure is self-consistent.

\begin{figure}[t!]
\renewcommand{\baselinestretch}{1.0}
\includegraphics[angle=0,width=.8\linewidth]{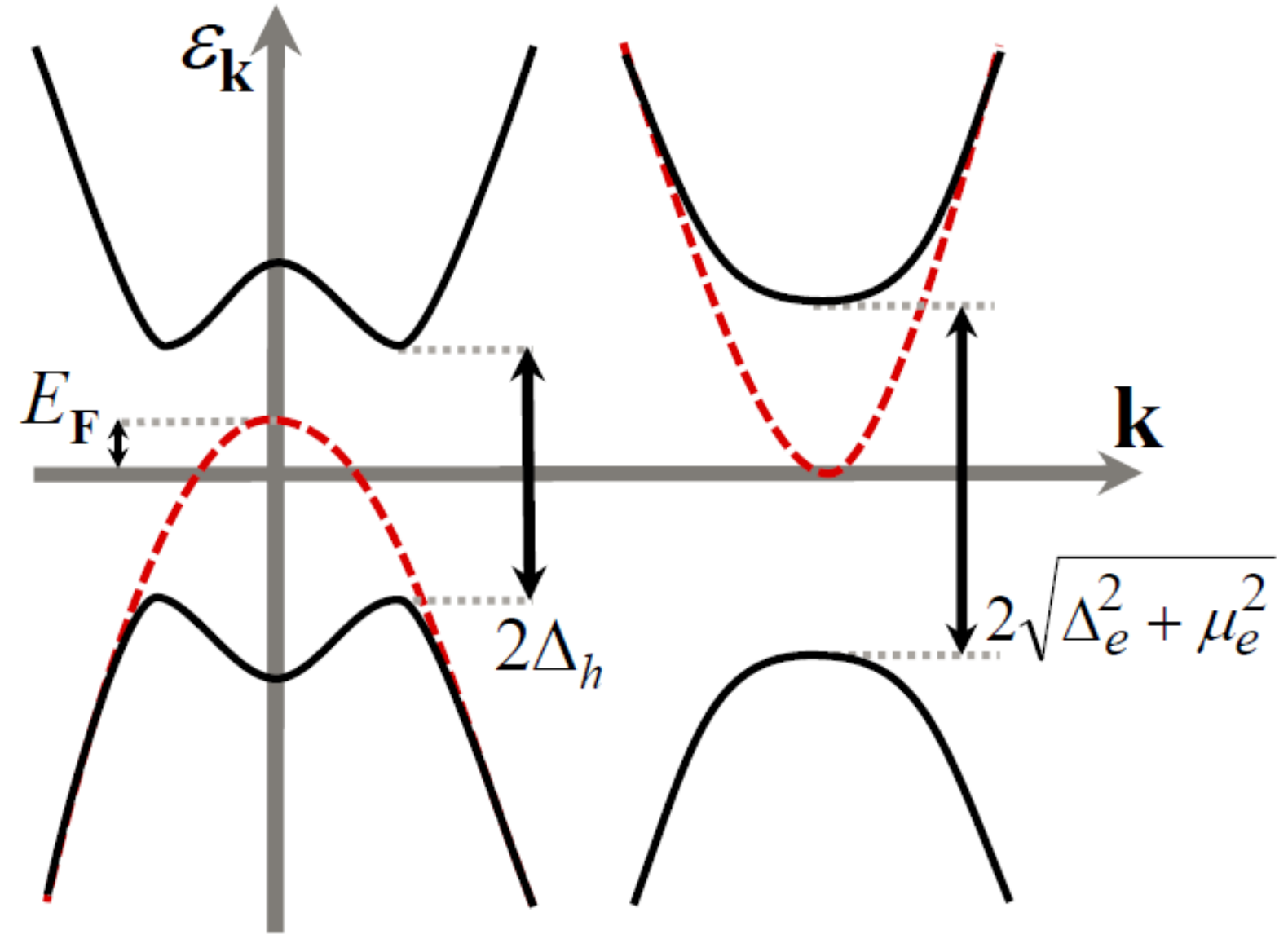}
\caption{Fermionic dispersions for the two-band model with one hole and one electron band in the limit $E_F \ll E_0$.
  Red dashed line -- the bare dispersions (the one which the system would have at $T=0$ in the absence of the pairing). Black lines -- the dispersions right above $T_{ins}$, blue lines -- the dispersion below $T_{ins}$.  The chemical potential for the electron band is negative, and the minimal gap is $\sqrt{\Delta^2_e + \mu^2}$.
   For the hole band, the chemical potential is positive and the dispersion $\pm \sqrt{\Delta^2_h + (k^2/(2m) - \mu_h)^2}$
   is non-monotonic, with the minimal energy $\Delta_h$ at $k_F = \sqrt{2m \mu_h}$.  Note that this $k_F$ is larger than the bare
 $k_{F,0} =\sqrt{2m E_F}$.   The dispersion at $E_F \geq E_0$ is quite similar.}
\label{fig:fig0_1}
\end{figure}

In Fig. \ref{fig:fig0_1} we plot the dispersions of fermions from hole and electron bands at $T > T_{ins}$ and $T \ll T_{ins}$ along with the
   bare dispersions (the one the system would have at $T=0$ in the absence of the pairing).  The figure is for the case   $E_F < E_0$,  the dispersion at $E_F > E_0$ is quite similat.   Observe that the minimal energy of a fermion
 on the hole band
(often associated with the "gap") is $\sqrt{\Delta^2_e + \mu^2_e}$, while the minimal energy of a fermion  on the electron band is just $|\Delta_h| \approx |\Delta_e|$, i.e.,
 it is smaller.

Returning to Eq. (\ref{ch_16}), we notice that the  temperature $T_{ins}$ is only slightly higher than $T_0 =1.13 E_0$,
 at which the scattering amplitude for two particles in a vacuum diverges (to obtain $T_0$ one just has to set $E_F =0$ in Eq.(\ref{ch_16})).   Like in the one-band case, this  poses the question what is the actual $T_c$ in this situation, because the development of the two-particle bound state does not generally imply the breaking of U(1) gauge (phase) symmetry.
 To understand what $T_c$ is we need to compute superconducting stiffness. This is what we do next.

\subsection{Superconducting $T_c$}
\label{sec_3d}

We express $U(1)$ order parameters $\Delta_e$ and $\Delta_h$ as $\Delta e^{-\phi_h}$ and $\Delta e^{-\phi_e}$ (we recall that, for equal masses $m_h=m_e$, the magnitudes of $\Delta_e$ and $\Delta_h$ are equal, up to small corrections). In the equilibrium $\phi_e-\phi_h=\pi$. To obtain the superfluid stiffness  at $T=0$ we need two ingredients~\cite{lara,ukr}:  the gradient  terms in the energy
 $(\nabla \phi_h)^2$,  $(\nabla \phi_e)^2$, and $(\nabla \phi_h) (\nabla \phi_e)$, and  the mixing term
 $\Delta_e \Delta^*_h + \Delta^*_e \Delta_h \propto \cos{(\phi_e - \phi_h)}$. The last term is important when the stiffnesses on the hole and the electron bands  substantially differ in magnitude because it generates the mass for phase fluctuations on the band with a smaller stiffness  once the phase of the gap on the band with a larger
  stiffness  gets ordered.
\begin{figure}[t!]
\renewcommand{\baselinestretch}{1.0}
\includegraphics[angle=0,width=1.0\linewidth]{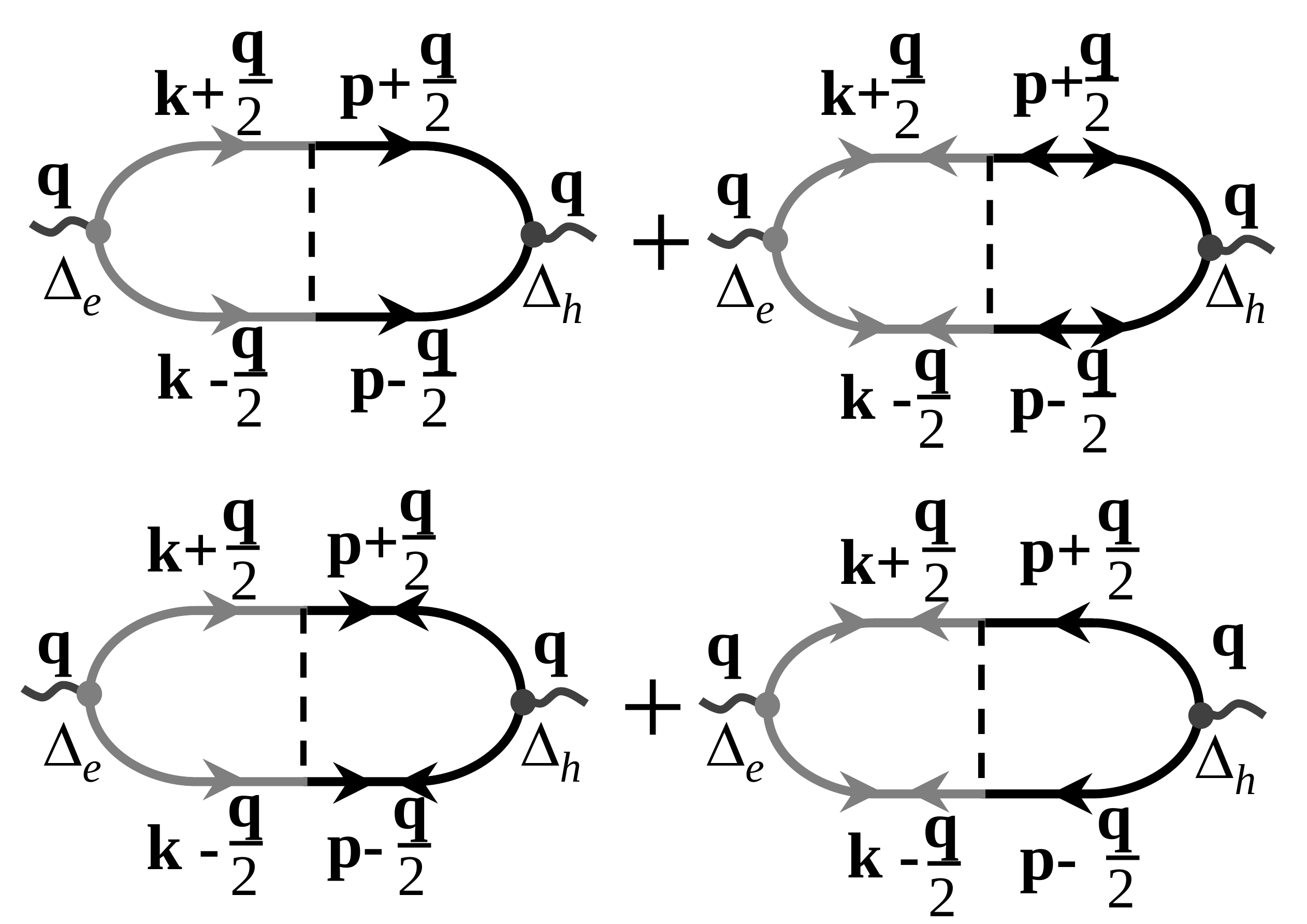}
\caption{Diagrammatic representation of the  two-loop diagram for the potential energy $E_{pot} (q)$ in two-band models with a constant inter-band interaction $U$ (the dashed line).
 Dark and light lines represent fermions from two different bands.  The prefactor for $q^2$ term in $E_{pot} (q)$ determines the superfluid stiffness.  For a constant (i.e., angle-independent) $U$,  the $q^2$ term appear by expanding $GG$ or $FF$ terms either on the right side of on the left side of each diagram, but there is no cross-term from taking linear in $q$ terms on the right and on the left. }
\label{fig4}
\end{figure}

The prefactors for the gradient terms can be evaluated in the same way as in the one-band model, by allowing the  phases to vary as $e^{i (\nabla \phi) r}$, i.e., by taking the Fourier transform $\Delta (q) = \Delta \delta (q-\nabla \phi)$, and evaluating the prefactors for the $q^2$ terms in the potential energy (see Sec. \ref{sec_2}).
   The cross term $\nabla \phi_h \nabla \phi_e$  could potentially come from the two-loop diagram shown in Fig.\ref{fig4}.  There are no symmetry restrictions which would prevent the cross term to be present~\cite{saurabh}, however, in our case the prefactor for $\nabla \phi_h \nabla \phi_e$  contains the integral $\int  d{\bf k} d{\bf p} ({\bf q} {\bf k}) ({\bf q} {\bf p}) U(k-p)$ (${\bf k}$ is near an electron band, ${\bf p}$ is near the hole band), which vanishes because we set $U(k-p)$ to be independent on the angle between ${\bf k}$ and ${\bf p}$. The gradient part of the energy  then comes solely from the bubbles made by fermions from the same band and is given by
   \beq
   \delta E_{gr} = \frac{1}{2} \rho_{s}^{h} (\nabla \phi_h)^2 + \frac{1}{2} \rho_{s}^{e} (\nabla \phi_e)^2
   \label{u_1}
   \eeq

   The two-loop diagram, shown in Fig.\ref{fig4}, however, gives rise to the mixed term $\Delta_e \Delta^*_h + \Delta^*_e \Delta_h$.  To see this we evaluate this diagram at $q=0$.
   The sum of $G_s(k)G_s (-k)$ and $F_s (k) F_s (-k)$ terms in the right and in the left gives exactly $1/U$, hence
   \beq
   \delta E_{mix} = \frac{1}{U} \left(\Delta_e \Delta^*_h + \Delta^*_e \Delta_h\right) = \frac{\Delta^2}{U} \cos{(\phi_e - \phi_h)}
   \label{u_2}
   \eeq
   Without loss of generality we can assume that in equilibrium $\phi_e =0$, $\phi_h =\pi$. Expanding in (\ref{u_2}) to quadratic order in deviations of $\phi_{e,h}$ from the equilibrium values and combining (\ref{u_1}) and (\ref{u_2}), we obtain fluctuation part of the energy in the form
   \beq
   \delta E_{fl} = \frac{1}{2} \left[\rho_{s}^{h} (\nabla {\tilde \phi}_h)^2 + \rho_{s}^{e} (\nabla {\tilde\phi}_e)^2 + \frac{\Delta^2}{U} \left({\tilde \phi}_e - {\tilde \phi}_h\right)^2 \right],
   \label{u_3}
   \eeq
   where ${\tilde \phi}_{e,h}$ are deviations from equilibrium values.   This $\delta E_{fl}$  can be treated as an effective Hamiltonian for fluctuations of ${\tilde \phi}$ in the sense that $\langle{\tilde \phi}^2\rangle \propto \int d {\tilde \phi} e^{-\delta E_{fl}/T}$.  This effective Hamiltonian can be also obtained by starting with fermionic Hamiltonian with 4-fermion interaction in the Cooper channel and using Hubbard-Stratonovich transformation to re-express the partition function as the integral over bosonic variables
   $\Delta_{e,h} (r) = |\Delta| e^{i\phi_{e,h} (r)}$ (see Refs.\cite{lara,ukr,lara_2,rafael_12,yuxuan}).

Transforming into the momentum space and evaluating $\langle |{\tilde \phi}_{e,h} (k)|^2 \rangle$, we obtain
\bea
\langle|{\tilde \phi}_{e} (k)|^2\rangle &=& \frac{T \left(k^2 \rho_{s}^{h} + \frac{\Delta^2}{U}\right)}{k^4 \rho_{s}^{e} \rho_{s}^{h} + \frac{\Delta^2}{U} k^2 \left(\rho_{s}^{e} + \rho_{s}^{h}\right)} \nonumber \\
 \langle|{\tilde \phi}_{h} (k)|^2\rangle &=& \frac{T \left(k^2 \rho_{s}^{e} + \frac{\Delta^2}{U}\right)}{k^4 \rho_{s}^{e} \rho_{s}^{h} + \frac{\Delta^2}{U} k^2 \left(\rho_{s}^{e} + \rho_{s}^{h}\right)}
\label{u_4}
\eea
If the mixing term was absent (i.e., if there was no $\Delta^2/U$ term in (\ref{u_4})), we would have $\langle|{\tilde \phi}_{e} (k)|^2\rangle = T/(\rho_{s}^{e} k^2)$ and $\langle|{\tilde \phi}_{h} (k)|^2\rangle = T/(\rho_{s}^{h} k^2)$, i.e., phase fluctuations of $\Delta_e$ and $\Delta_h$ would be decoupled and the actual $T_c$, below which the system displays full coherence, would be determined by the smaller  of the two stiffnesses.  The presence of the mixed term changes the situation because now the ordering of one ${\tilde \phi}$ produces the mass term for fluctuations of the  other phase variable.  To estimate $T_c$ we need to look at the small momentum asymptotic of Eq.(\ref{u_4}) where
\beq
\langle|{\tilde \phi}_{e} (k)|^2\rangle =  \langle|{\tilde \phi}_{h} (k)|^2\rangle  = \frac{T}{\left(\rho_{s}^{h} + \rho_{s}^{e}\right) k^2}.
\label{u_5}
\eeq
 Hence $T_c$ is determined by the combined stiffness $\rho_{comb} = \rho_{s}^{e} + \rho_{s}^{h}$.   When $\rho_{comb} \gg T_{ins}$,  phase fluctuations are costly and
 $T_c$ almost coincides with $T_{ins}$. When $\rho_{comb} \ll T_{ins}$, we  again use the approximate criterium
 $T_c \approx  (\pi/2) \rho_{comb} (T) =1.57  \rho_{comb} (0)$.
 Note in passing that the combined stiffness $\rho_{comb}$ also appears in the dispersion of the anti-phase (Leggett) mode of phase oscillations in a two-band superconductor~\cite{ukr}.
  We now proceed with the calculations of $\rho_{s}^{e}$ and $\rho_{s}^{h}$.
   The $q-$dependent part of the energy for each band is again given by the sum of the convolutions of normal and anomalous Green's functions with the total momentum $q$, with $\Delta$ in the vertices, i.e., by the integrals of $\Delta^2 (G_s (k+q/2) G_s (-k+q/2) + F_s (k+q/2) F_s (-k+q/2))$ (see Eq. ( \ref{ch_e8})).
     Each stiffness at $T=0$  is then given by
    Eq. (\ref{ch_e10}) with $\mu = \mu_e$ or $\mu_h$, i.e.,
    \bea
    \rho_{s}^{e} (T=0) &=& \frac{\Delta^2_e}{8\pi} \int_0^\Lambda d \varepsilon \frac{\varepsilon}{((\varepsilon - \mu_e)^2 + \Delta^2_e)^{3/2}} \nonumber \\
    \rho_{s}^{h} (T=0) &=& \frac{\Delta^2_h}{8\pi} \int_0^\Lambda d \varepsilon \frac{\varepsilon}{((\varepsilon - \mu_h)^2 + \Delta^2_h)^{3/2}}.
    \label{nn_2}
    \eea
    Using the same manipulations as in Sec. \ref{sec:sc_one_band}, one can relate $\rho_{s}^{h}$ and $\rho_{s}^{e}$ to the total number of fermions in the hole and the electron band, $N_h$ and $N_e$. The two are  given by
    \bea
    N_e &=& N_0 \int_0^\Lambda d \varepsilon \left(1 -\frac{\varepsilon - \mu_e}{\sqrt{(\varepsilon - \mu_e)^2 + \Delta^2_e}}\right) \nonumber \\
     N_h &=& N_0 \int_0^\Lambda d \varepsilon \left(1 +\frac{\varepsilon - \mu_h}{\sqrt{(\varepsilon - \mu_h)^2 + \Delta^2_h}}\right),
    \eea
    and the relations are
    \beq
    \frac{N_e}{N_0} = 8\pi \rho_{s}^{e},~~ \frac{N_h}{N_0} = 2\Lambda -8\pi \rho_{s}^{h}
    \eeq
    The conservation of the total number of particles implies that $N_h + N_e = 2 N_0 (\Lambda - E_F)$, hence
    \beq
     \rho_{s}^{h} - \rho_{s}^{e} = \frac{E_F}{4\pi}
     \label{nn_1}
     \eeq
     This condition, however, only specifies the difference between $\rho_{s}^{h}$ and $\rho_{s}^{e}$, the combined stiffness $\rho_{comb}$ is not fixed and depends on the ratio $E_F/E_0$, as we see below.

      At $E_F \gg E_0$ we obtained from Eq.(\ref{nn_2}), using $\Delta \ll E_F$ and $\mu_e (T=0) = -\frac{\Delta}{2\sqrt{2}}, ~~\mu_h (T=0) = E_F - \mu_e (T=0) \approx E_F$:
   \bea
 \rho_{s}^{h} (T=0) & \approx &  \frac{E_F}{4\pi}, \nonumber\\
 \rho_{s}^{e} (T=0) & \approx  & 0.71 \frac{\Delta_h}{8\pi} =0.05 T_{ins} \ll \rho_{s}^{h} (T=0),
   \eea
   where, we remind, $T_{ins} \sim \Delta \sim (E_F E^2_0)^{1/3} \ll E_F$.
  Adding the two stiffnesses, we find $\rho_{comb} (T=0) \approx E_F/(4\pi) \gg T_{ins}$. The inequality $\rho_{comb} \gg T_{ins}$ implies that
   phase fluctuations are costly and hence $T_c \approx T_{ins}$.

  Note in passing that at $E_F \gg E_0$,   $T/(\rho_{comb} k^2)$ behavior of the the Fourier transform of the correlation function for phase fluctuations holds in the full momentum range where the
 gradient expansion is applicable.  Indeed, gradient expansion holds when $k$ is smaller that inverse superconducting coherence length $\xi^{-1} = \Delta/v_F$.
  Comparing the mixed term $\Delta^2/U$ with even the larger $k^2 \rho_{s}^{h} \sim k^2 E_F$, we find that at $k \sim \xi^{-1}$,
  $k^2 \rho_{s}^{h} \sim m \Delta^2$ is already parametrically smaller  than $\Delta^2/U \sim (m \Delta^2) /\lambda$. Hence Eq. (\ref{u_4}) is valid for all $k < \xi^{-1}$.

At $E_F= E_0$  we obtain from Eq.(\ref{nn_2}), using the results for $\Delta_{e,h}$ and $\mu_{e,h}$ from Sec. \ref{sec_3c},
 \beq
 \rho_{s}^{h} (T=0) = 2.05 \frac{E_F}{4\pi},~~ \rho_{s}^{e} (T=0) = 1.05 \frac{E_F}{4\pi}
 \eeq
 Observe that $\rho_{s}^{h} (T=0)- \rho_{s}^{e} (T=0) = E_F/(4\pi)$, as it should be, according to (\ref{nn_1}).   Combining the two we obtain
  $\rho_{comb} = 3.1 E_F/(4\pi)$. Using $T_c \approx 1.57 \rho_{comb}$ we obtain $T_c \approx 0.39 E_F$.  The onset temperature for the pair formation is $T_{ins} = 1.35 E_F \approx 3.49 T_c$. We see that now $T_{ins}$ is substantially larger than $T_c$, hence already at $E_F = E_0$ the system should display
   preformed pair behavior  in a wide range of temperatures.

 Finally, at $E_F \ll E_0$ the chemical potentials $\mu_{e,h} \sim E_F$ are parametrically smaller than $\Delta$.  In this situation we obtain from Eq.(\ref{nn_2})
 \beq
 \rho_{s}^{h} \approx \rho_{s}^{e} \approx  \frac{\Delta^2}{8\pi} \int_0^\infty d\varepsilon \frac{\varepsilon}{(\varepsilon^2+\Delta^2)^{3/2}} =
 \frac{\Delta}{8\pi} = 0.07 T_{ins}.
 \label{ch_e11}
 \eeq
 Hence
 \beq
 \rho_{comb} = 0.14 T_{ins},  ~~T_c \approx 0.22 T_{ins}
 \eeq
 This holds even when $E_F=0$, i.e., when the electron band is empty.  The stiffnesses $\rho_{s}^{h}$ and $\rho_{s}^{e}$ are equal in this limit, as required by (\ref{nn_1}) but each remains non-zero and of order $T_{ins}$.  As the consequence, $T_c$ remains finite and  also of order $T_{ins}$. Still, because
   numerically  $T_c \ll T_{ins}$,  there exists a sizable temperature range of preformed pair behavior.

     Because now $T_{ins}$
   almost coincides with the temperature $T_0$ of bound state formation for two particles in a vacuum, the change of system behavior between $E_F \gg E_0$ and $E_F \ll E_0$  can be interpreted as BEC phenomenon.
    We emphasize, however, that the ratio of $\frac{T_{ins}}{T_c}$ remains finite at $E_F \to 0$, in distinction to ordinary BCS-BEC crossover, where this ratio tends to infinity when $E_F$ vanishes.~\cite{Mohit89}

The still existence of a finite $T_c$ at vanishing $E_F$ is in variance with the situation in the one-band model and, as we will see in the next Section, also with
  the two-band model with two electron/two hole bands. There, $T_c$ vanishes when
 $E_F=0$ on both bands.   The difference can be easily understood because in the other two models there are no carriers at $E_F =0$  to form superconducting condensate, hence the gap must vanish at $E_F=0$, otherwise there would appear an image band at negative energies with a finite density of carriers in it.
  In the model with a hole and an electron band there is charge reservoir in the hole band, and the charge density can be re-distributed into the image bands even at $E_F=0$.
   The image of the electron band appears at negative energies $E = -\sqrt{\Delta^2_e + (\varepsilon^e_k-\mu_e)^2}$.  The states in this new band are filled by electrons, and their total density is given by $N_0 \int d\varepsilon_k \left(1 - \frac{\varepsilon^e_k -\mu_e}{\sqrt{\Delta^2_e + (\varepsilon^e_k-\mu_e)^2}}\right) \sim \Delta \sim T_{ins}$.  The electrons from the filled states in this image band can form superconducting condensate, and, because all energy scales are of order $T_{ins}$, superconducting $T_{c}$ is also a fraction of $T_{ins}$.

Note also that at $E_F \ll E_0$, the Fourier transform of the correlation function for phase fluctuations is given by $T/(\rho_{comb} k^2) = T/(2 \rho_{s}^{h} k^2) $  at the lowest $k$, but crosses over to a similar but not identical expression at larger $k$, which are still smaller than $\xi^{-1}$. The reasoning is that $k^2 \rho_{s}^{h} \approx k^2 \rho_{s}^{e}$ becomes comparabvle to the mixing term $\Delta^2/U$ at $k_{typ} \sim \xi^{-1} (E_F/\lambda E_0)^{1/2}$, which, at small enough $E_F$, is smaller than $\xi^{-1}$. In between $k_{typ}$ and $\xi^{-1}$,  the correlation function scales as $T/(\rho_{s}^{h} k^2)$, i.e., the functional form is the same as at the smallest $k$ but the prefactor differs by $2$.

\subsection{The density of states at $T=0$}

\begin{figure}[t!]
$\begin{array}{c}
\renewcommand{\baselinestretch}{1.0}
\includegraphics[angle=0,width=1.0\linewidth]{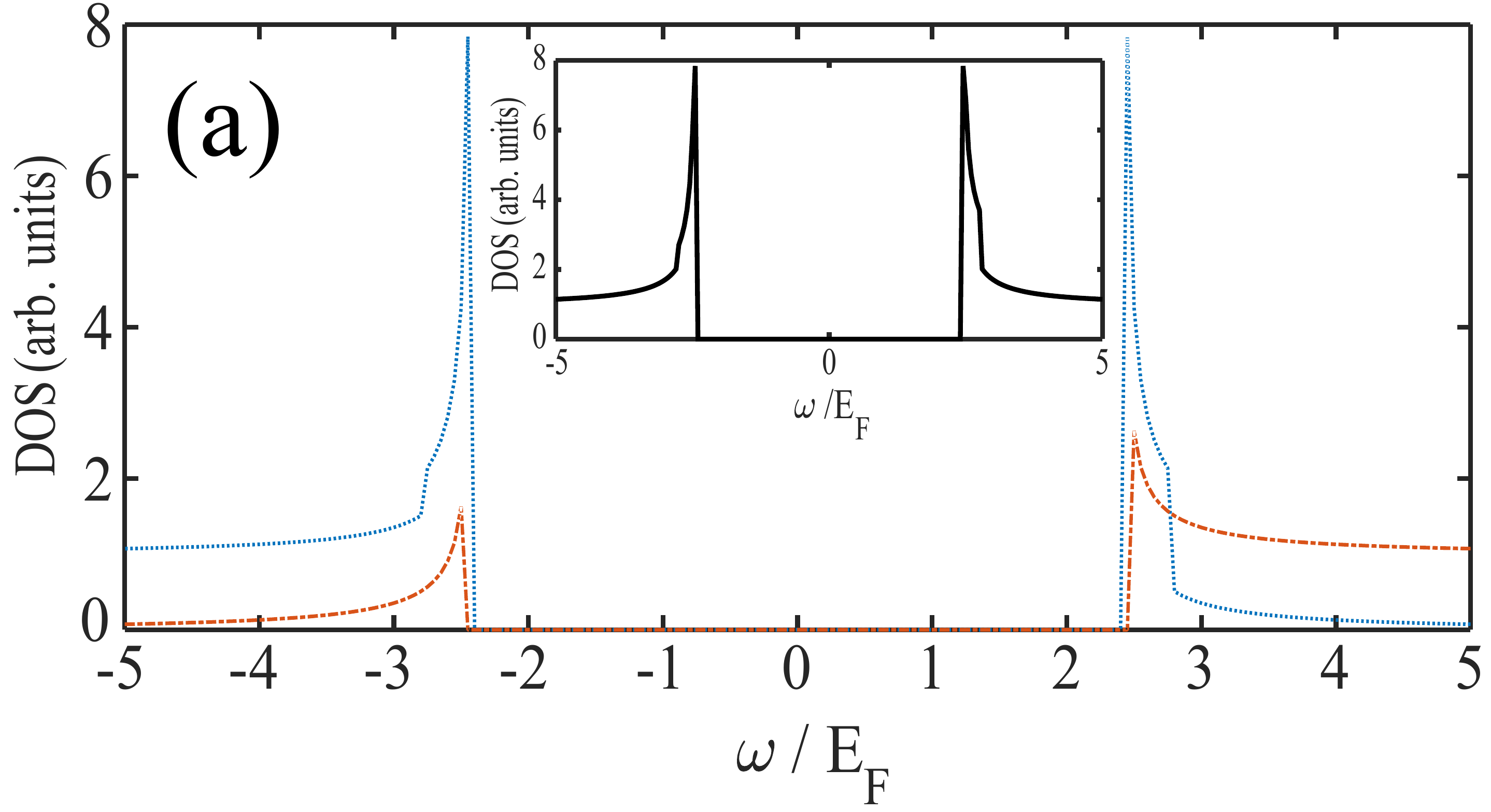}\\
\includegraphics[angle=0,width=1.0\linewidth]{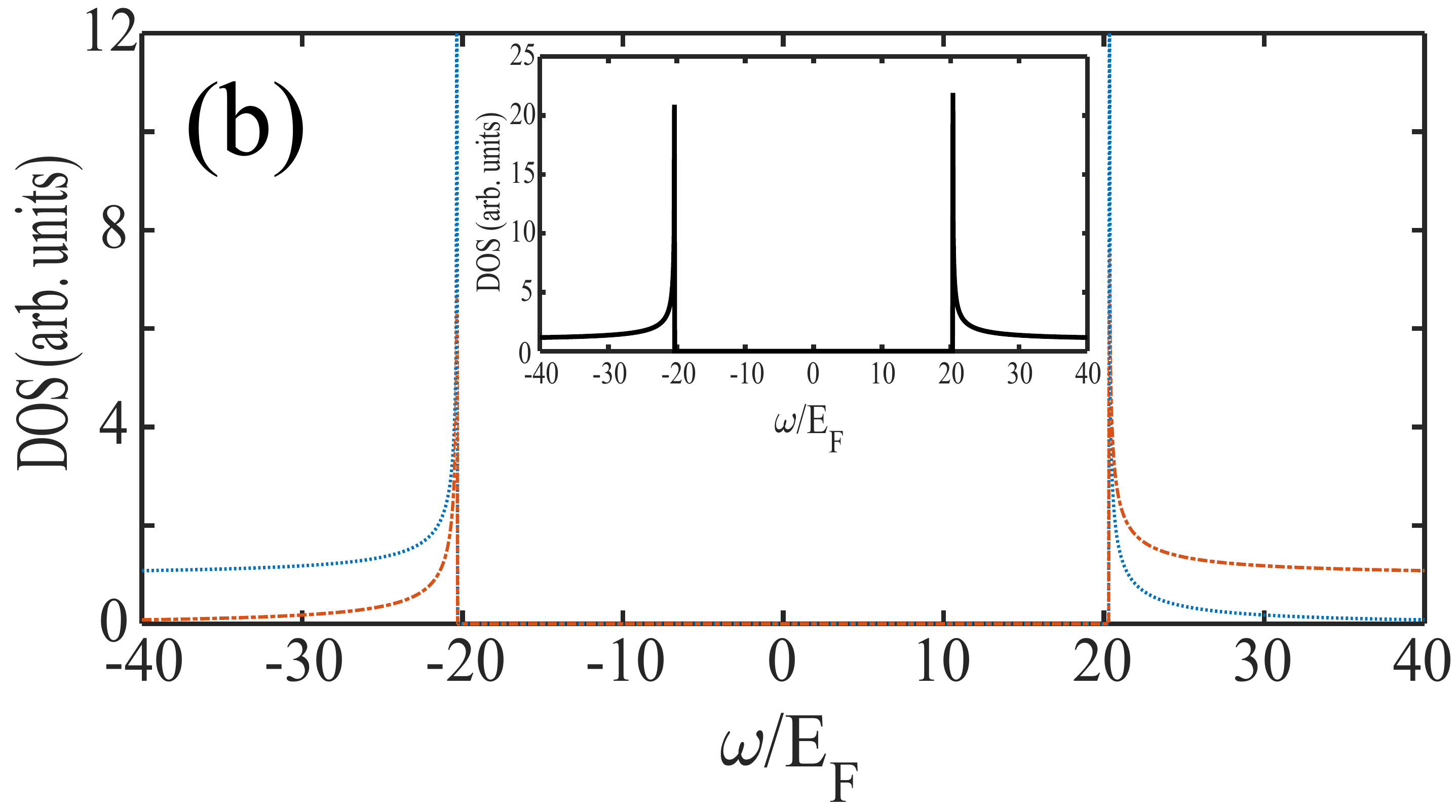}
\end{array}$
\caption{The DOS at $T=0$ for the model with one hole and one electron pocket. Panel (a) -- $E_0 = E_F$,  panel (b) -- $E_0 =10 E_B$.
  For $E_0 = E_F$, $\mu_e = -0.34 E_F, \mu_h = 1.34 E_F$, and $\Delta_h \approx - \Delta_e = 2.43 E_F$.  For $E_F=0.1E_0$,
  $\mu_h \approx \frac{3E_F}{2}$, $\mu_e \approx -\frac{E_F}{2}$, and $\Delta_e \approx -\Delta_h \approx 20.3 E_F$.  To cut the singularities and make other features of DOS visible, we added fermionic damping $\gamma=0.001E_F$.
  Main figures -- the DOS separately for the hole band (dashed blue) and the electron band (dashed red).  For the hole band, the DOS has a  square-root singularity at  $|\omega| =|\Delta_h|+0$, symmetric between negative and positive frequencies, and  van-Hove discontinuity at $|\omega| = \sqrt{\Delta_h^2+\mu_h^2}+0$ (see Eq. (\protect\ref{nn_a}); for hole dispersion positive and negative frequencies in (\ref{nn_a}) have to be interchanged). The latter is stronger for positive frequencies, due to  anisotropy of coherence factors.  For $E_F = 0.1 E_0$, the singularity and the discontinuity are almost undistinguishable.
  For the electron band,  the DOS jumps to finite value at $|\omega| =\sqrt{\Delta_e^2+\mu_e^2} +0$ and
   is highly anisotropic between negative and positive frequencies, again due to  anisotropy of the coherence factors (see Eq. (\protect\ref{nn_b})).
    For $E_F=0.1 E_0$, $\Delta >> \mu_{e}$, and the
   DOS right after the jumps  are large, of order $\Delta/\mu_{e}$.
 Insets -- the total DOS. Observe that at large frequencies the total DOS tends to a finite value for  both positive and negative $\omega$.  }
\label{fig_twoband}
\end{figure}

The DOS in the model with one hole and one electron band is different from that in the one-band model because now fermionic excitations in the normal state exist at both positive and negative frequencies.  In the main parts of the two panels in Fig.\ref{fig_twoband} we show the behavior of the DOS at $T=0$ separately for hole and electron bands, $E_0 = E_F$ and $E_0 \gg E_F$, respectively  The behavior
 of the DOS on the electron band is very similar to that in the one-band model (see Eq. (\ref{nn_b}) and Fig.6).  Namely, the DOS vanishes at $|\omega| < \sqrt{\Delta_e^2+\mu_e^2}$ and
  jumps to a finite value at $|\omega| =\sqrt{\Delta_e^2+\mu_e^2} +0$. The DOS is highly anisotropic between negative and positive frequencies due to anisotropy of
   coherence factors. It is much larger at positive frequencies, where it tends to a finite value at large $\omega$. At negative frequencies, the discontinuity is
    weaker (and rapidly suppressed by a fermionic damping), and the DOS falls off as $1/\omega^2$ for larger negative frequencies.  On the hole band, the DOS vanishes
     at $|\omega| < |\Delta_h|$ and has a BCS-like square-root singularity at  $|\omega| =|\Delta_h|+0$, symmetric between negative and positive frequencies.  At larger $|\omega|$, the DOS on the hole band has a discontinuity at $|\omega| = \sqrt{\Delta_h^2+\mu_h^2}-0$, when $|\omega|$ crosses the edge of the band and the corresponding momentum $k=0$ (see Eq. (\ref{nn_a}); for hole dispersion positive and negative frequencies in (\ref{nn_a}) have to be interchanged).
     In the normal state  this would be van-Hove discontinuity at the top of the hole band. In a superconductor, the discontinuity holds for both positive and negative $\omega$, but the coherence factor is much larger for a positive $\omega$. At higher frequencies, the DOS on the hole band tends to a finite value at negative frequencies and scales as $1/\omega^2$ at positive frequencies.

     In the insets of Fig.\ref{fig_twoband} we show the total (combined) DOS for $E_0 = E_F$ and $E_0 \gg E_F$.
      Observe that DOS tends to a finite value at both positive and negative frequencies.

\section{Two-band model with two electron bands}
\label{sec_3_n}

In this section we analyze the model with two bands of equal type. The results are identical for the model with two hole bands and for the one with two electron bands.
 For definiteness we consider the model with two electron bands as in Fig.\ref{fig1}(c).   We consider the same electronic configuration at $T=0$ as  for one hole/one electron band model. Namely, we set the chemical potential to touch the bottom of one of the bands
  and cross the other band, i.e. the Fermi energy to zero in one band (band 2) and  finite in the other (band 1).  At the end of this Section we consider how the onset temperature for the pairing evolves  when we move the chemical potential such that both bands cross the Fermi level. Like in the previous section, we restrict with inter-band (pair-hopping) pairing interaction.

 The behavior of $T_{ins}, \Delta$, $\mu$, and $\rho_s$  in the model with two electron bands (and in a more general model with intra-band pairing interaction)
 has been discussed analytically in Refs. \cite{Varlamov2000} for
  the case when $E_F \gg E_0$.  The set of equations for $T_{ins}, \Delta$, and $\mu$ has been solved numerically for arbitrary $E_F/E_0$
    (Refs.~\cite{DeMelo2006,Guidini2014}). When the comparison is possible, our results agree with these works, but we also present new analytical results for $T_{ins}, \Delta$, $\mu$,
     and the superfluid stiffness for the cases when $E_F \sim E_0$ and $E_F < E_0$.
      The pairing at $E_F \gg E_0$  has been considered recently in Refs.~\cite{Fernandes2013,Innocenti2010}. Our results agree with these earlier works modulo that  they computed $T_{ins}$ without  including into consideration the temperature dependence of the chemical potential in the band 2, while we argue that this renormalization is $O(1)$ effect.

  The analysis of the bound state energy for two particles at $E_F \equiv 0$ does not differ from that in previous Sections, and the result is that  the scattering amplitude diverges at $T_0 = 1.13 E_0$, where, like before,  $E_0 = \Lambda e^{-2/\lambda}$.  The bound state energy at $T=0$ is $2 E_0$.

The onset temperature for the pairing at a finite $E_F$  is obtained by solving simultaneously the linearized  equations for $\Delta_1$ and $\Delta_2$ and the equations for the chemical potentials $\mu_1 (T)$ and $\mu_2 (T)$, subject, in this case,  to $\mu_1 (T) -\mu_2 (T) = E_F$.   The set of equations is ($\mu_1 = \mu_1 (T_{ins}), \mu_2 = \mu_2 (T_{ins})$):
 \bea
\Delta_1 &=& - \frac{\lambda}{2} \Delta_2 \int_{-\mu_2}^{\Lambda}  \frac{dx}{x} \tanh{\frac{x}{2T_{ins}}} \nonumber \\
\Delta_2 &=& -  \frac{\lambda}{2} \Delta_1 \int_{-\mu_1}^{\Lambda}  \frac{dx}{x} \tanh{\frac{x}{2T_{ins}}} \nonumber \\
E_F &=& T \log{\left[\left(1 + e^{\mu_1/T}\right)\times \left(1 + e^{\mu_2/T}\right)\right]} \nonumber \\
\mu_1 &-& \mu_2 = E_F
\label{ch_11_n}
\eea
The first two equations reduce to
\beq
\frac{4}{\lambda^2} =  \int_{-\mu_1}^{\Lambda}  \frac{dx}{x} \tanh{\frac{x}{2T_{ins}}} \times \int_{-\mu_2}^{\Lambda}  \frac{dx}{x} \tanh{\frac{x}{2T_{ins}}}
\label{ch_11_0_n}
\eeq

 Below $T_{ins}$,  $\Delta_1$ and $\Delta_2$ become non-zero and one has to consider non-linear gap equations and modify the equation for the chemical potential.
 At $T=0$ we have [$\mu_1 = \mu_1 (T=0), \mu_2 = \mu_2 (T=0)$]:
\bea
&&\Delta_1 = - \frac{\lambda}{2}  \Delta_2 \log{\frac{2\Lambda}{\sqrt{\mu^2_2+ \Delta^2_2} - \mu_2}}\nonumber \\
&&\Delta_2 = - \frac{\lambda}{2}  \Delta_1 \log{\frac{2\Lambda}{\sqrt{\mu^2_1+ \Delta^2_1} - \mu_1}} \nonumber \\
&& 2 E_F = \sqrt{\mu^2_1+ \Delta^2_1} + \sqrt{\mu^2_2 + \Delta^2_2} + \mu_1 + \mu_2  \nonumber \\
&& \mu_1 - \mu_2 = E_F
\label{ch_11_1_n}
\eea

 \subsection{The case $E_F  \gg E_0$}
\label{sec_3na}

Like we did in the previous Section, we  assume and then verify that  $T_{ins} \ll E_F$ and that at $T =T_{ins}$, $\mu_1 \approx E_F$, while $\mu_2 \sim T_{ins}$
   Under these assumptions, the equations on the chemical potentials in (\ref{ch_11_n}) yield ($\mu_{1,2} \equiv \mu_{1,2} (T_{ins})$)
\beq
\mu_2 = T \log{\left(\frac{1}{1 + e^{\mu_2}{T}}\right)}, ~~ \mu_1  = E_F + \mu_2
\eeq
Solving for $\mu_2$ we obtain
\beq
\mu_2 = T_{ins} \log{\frac{\sqrt{5}-1}{2}} = -0.48 T_{ins}, ~~\mu_1  = E_F -0.48 T_{ins}
\label{fri_4_n}
\eeq
This is the same result as Eq. (\ref{fri_4}).
Substituting these results into the first two equations in (\ref{ch_11}) we obtain after simple algebra
\bea
\Delta_2 &=& - \frac{\lambda}{2}  \Delta_1 \left[2\log{\frac{1.13 \Lambda}{T_{ins}}} - \frac{2}{\lambda} + \log{\frac{E_F}{E_0}}\right] \nonumber \\
\Delta_1 &=& -  \frac{\lambda}{2}  \Delta_2 \log{\frac{1.13 D \Lambda}{T_{ins}}}
\label{ch_11_2_n}
\eea
 where $D = 0.79$.
Solving the set we obtain
\beq
T_{ins} = 1.13 D^{1/3} E^{1/3}_F E^{2/3}_0 = 1.04 E^{1/3}_F E^{2/3}_0
\label{ch_12_n}
\eeq
in full similarity with Eq. (\ref{ch_12}) for the model with a hole and an electron pocket.
Like in that case, Eq. (\ref{ch_12_n}) is valid when $\lambda \log{E_F/E_0} \ll 1$.  At even larger $E_F \leq \Lambda$, when $\lambda \log{E_F/E_0} = O(1)$,  $T_{ins}$ is given by
\beq
T_{ins} \sim \Lambda \left(\frac{E_F}{\Lambda}\right)^{1/4} e^{-\frac{\sqrt{2}}{\lambda}}
\label{ch_14_n}
\eeq

The ratio of the gaps on electron and hole bands at $T = T_{ins} -0$ is
\beq
\frac{\Delta_2}{\Delta_1} \approx -\left(1 + \frac{\lambda}{6} \log{\frac{E_F}{E_0}}\right)
\label{ch_15_1_n}
\eeq
We see that the gap on the band which touches the Fermi level
is {\it larger} than the gap on the band, which crosses Fermi level, in full agreement with the case of one hole and one electron band.
 This result holds even when  the full band 2 is located above the Fermi level.  The ratio of $\Delta_2/\Delta_1$ increases when $E_F$ gets larger  and approaches $\sqrt{2}$ when $E_F$ becomes of order $\Lambda$.

At $T=0$,  the solution of the set (\ref{ch_11_1_n}) at $E_F \gg E_0$ but $\lambda \log{E_F/E_0} \ll 1$ shows that the
 ratio of $\Delta_1$ and $\Delta_2$ remains the same as in Eq.  (\ref{ch_15_1_n}), i.e., up to subleading terms
 $\Delta_1 (T=0) = -\Delta_2 (T=0) = \Delta$.
 Solving for the chemical potentials we then find ($\mu_{1,2} \equiv \mu_{1,2} (T=0)$)
 \beq
 \mu_2  = -\frac{\Delta}{2\sqrt{2}}, ~~\mu_1  = E_F  + \mu_2
 \eeq
 Substituting this into the first two equations in Eq. (\ref{ch_11_1_n}) and solving for $\Delta$ we obtain
 \beq
 \Delta = 2^{5/6} E^{1/3}_F E^{2/3}_0 = 1.78 E^{1/3}_F E^{2/3}_0
 \eeq

\subsection{The case $E_F = E_0$}
 \label{sec_3nc}

Like we did in the previous section, we express $T_{ins} = a E_F, \mu_1 (T_{ins}) = b E_F, \mu_2 (T_{ins})= E_F (b-1)$.
Substituting these relations into Eq.(\ref{ch_11_n}) and using the fact that
  \beq
  \frac{2}{\lambda} \int_0^\Lambda \frac{\tanh{\frac{x}{2T_{ins}}}}{x} = \frac{2}{\lambda} \log{1.13 \Lambda/T_{ins}} = 1 - \frac{2}{\lambda} \log{\frac{a}{1.13}},
  \eeq
 we obtain, to leading order in $\lambda$, the set of two equations on  $a$ and $b$:
\bea
&&1 = a \log{\left[\left(1 + e^{\frac{b-1}{a}}\right) \times \left(1 + e^{\frac{b}{a}}\right)\right]} \nonumber \\
&&2 \log{\frac{a}{1.13}} = \int_0^{\frac{b}{2a}} dy \frac{\tanh{y}}{y} + \int_0^{\frac{b-1}{2a}} dy \frac{\tanh{y}}{y}
\label{ch_16aa_n}
\eea
Solving the set we obtain $T_{ins} = 0.924 E_F, \mu_1  (T_{ins}) = 0.115 E_F$, and $\mu_2 (T_{ins}) = -0.885 E_F$.   As expected,
 the chemical potential $\mu_2$ becomes negative at a finite temperature.

At $T=0$ we express $\Delta = {\bar c} E_F, ~\mu_1 (T=0)={\bar b} E_F,~ \mu_2 (T=0)=({\bar b}-1) E_F$.
 Substituting into Eq. (\ref{ch_11_1_n}) we obtain the set of two equations
 \bea
 &&\left(\sqrt{{\bar b}^2+{\bar c}^2} - {\bar b}\right) \times \left(\sqrt{({\bar b}-1)^2 +{\bar c}^2} + (1-{\bar b})\right) =4 \nonumber \\
 && \sqrt{({\bar b}-1)^2 +{\bar c}^2} + \sqrt{{\bar b}^2+c^2} = 3-2{\bar b}
 \eea
 Solving this set we find ${\bar c} = 1.38$ and ${\bar b} = -0.05$, i.e., $\mu_1 (T=0) = -0.05 E_F, \mu_2 (T=0) = -1.05 E_F$, and $\Delta_1 \approx -\Delta_2 = 1.38 E_F$.
 We see that $\mu_1$  changes from a slightly positive to a slightly negative value  between $T= T_{ins}$ and $T=0$.

\subsection{The case $E_F \ll E_0$}
 \label{sec_3nb}

We assume and then verify that at small $E_F$ both $\mu_1$ and $\mu_2$ become negative at $T= T_{ins}$, and each exceeds $E_F$ by magnitude. Solving
 Eq. (\ref{ch_11_n}) under these assumptions we obtain
 \bea
 \mu_1 (T_{ins})  & = & - \frac{T_{ins}}{2} \log{\frac{T_{ins}}{E_F}} + \frac{E_F}{2}, \nonumber\\
 \mu_2 (T_{ins}) & = &  - \frac{T_{ins}}{2} \log{\frac{T_{ins}}{E_F}} - \frac{E_F}{2}
 \label{yyy_1}
 \eea
 Substituting these chemical potential into the equation for $T_{ins}$ an solving it, we obtain
 \beq
 T_{ins} = \frac{4.52 E_0}{\log{\frac{E_0}{E_F}}}
 \label{yyy_2}
 \eeq
 This $T_{ins}$ has the same functional form as $T_{ins}$ for the one-band model in the same limit $E_F \ll E_0$.  This $T_{ins}$ scales as $E_0$, up to a logarithmic factor, but still vanishes at $E_F=0$ due to logarithmic suppression.

 Plugging $T_{ins}$ from (\ref{yyy_2}) back into (\ref{yyy_1}) we obtain $\mu_1 \approx \mu_2 \approx -2.51 E_0$, what justifies the assumption we made.

Solving next for the gaps and chemical potentials at $T=0$,   we obtain (for $m_h = m_e$)  that $\mu_1 \approx \mu_2 \approx -E_0$, while
   $\Delta_1 \approx - \Delta_2 = \Delta$, where
 \beq
 \Delta = \sqrt{2} \sqrt{E_F E_0} \ll T_{ins}
 \label{yyy_3}
 \eeq
The expression for the gap also agrees, up to an overall factor, with that in the one-band model.

\begin{figure}[t!]
\renewcommand{\baselinestretch}{1.0}
\includegraphics[angle=0,width=.8\linewidth]{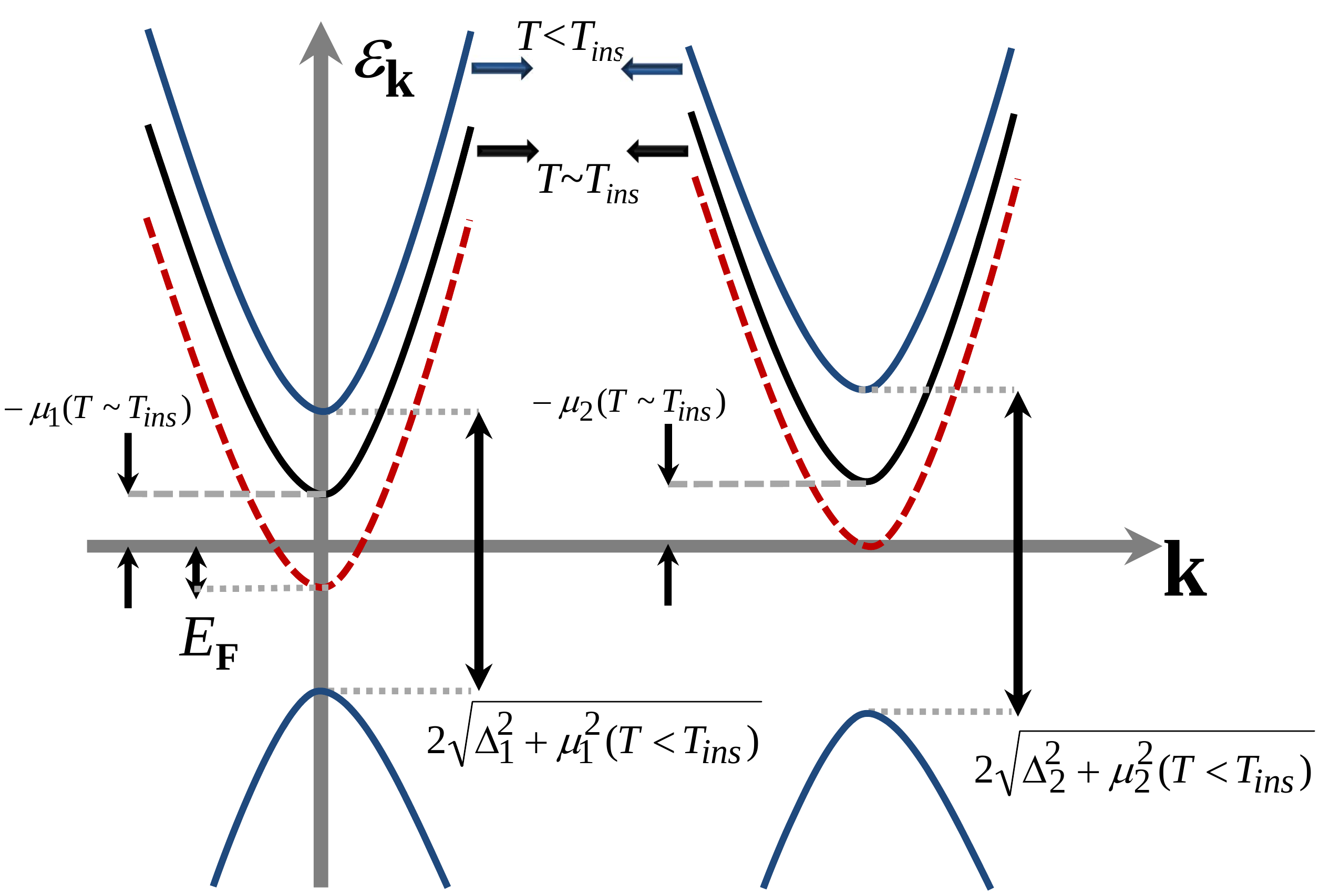}
\caption{Fermionic dispersions for the two-band model with two electron bands in the limit $E_F \ll E_0$.
  Red dashed line -- the bare dispersions (the one which the system would have at $T=0$ in the absence of the pairing). Black lines -- the dispersions right above $T_{ins}$, blue lines -- the dispersion below $T_{ins}$.  The chemical potentials for both bands are now negative at $T \geq T_{ins}$ and in the BEC state below $T_{ins}$, and the
    minimal gaps on each band is $\sqrt{\Delta^2 + \mu^2}$.  The minimum of dispersion on both bands is at $k=0$.  This is very similar to the behavior of the one-band model
     (see Fig. \ref{fig:bands_1}).}
\label{fig:two_el_bands}
\end{figure}

In Fig. \ref{fig:two_el_bands} we plot the actual dispersion of the two electron bands below $T_{ins}$ for the case $E_F  \ll E_0$, along with the bare dispersion.
 We see that the system behavior is very similar to that in the  one-band model (see Fig. \ref{fig:bands_1} for comparison).

To understand what $T_c$ is we again need to compute  superconducting stiffness.

\subsection{Superconducting $T_c$}
 \label{sec_3nd}

We follow the same logics as in the previous section, i.e., introduce coordinate-dependent phases of the gaps on the two electron bands
 $\phi_1 (r)$ and $\phi_2 (r)$, compute the prefactors for the gradient term in the ground state energy $E_{gr} = (1/2) \rho_{s,1} (\nabla \phi_1)^2 + (1/2) \rho_{s,2}
 (\nabla \phi_2)^2$ and the mixing term $(\Delta^2/U)  \cos{(\phi_1 - \phi_2)}$.   Performing the same calculations as in the previous Section, we find that
  $T_c$ is determined by the combined stiffness $\rho_{comb} = \rho_{s,1} + \rho_{s,2}$.

  The stiffnesses $\rho_{s,1}$ and $\rho_{s,2}$ are expressed via $\Delta_{1,2}$ and $\mu_{1,2}$ by the same formulas as we obtained in the previous two Sections:
    \bea
    \rho_{s,1} (T=0) &=& \frac{\Delta^2_1}{8\pi} \int_0^\Lambda d \varepsilon \frac{\varepsilon}{((\varepsilon - \mu_1)^2 + \Delta^2_1)^{3/2}} \nonumber \\
    \rho_{s,2} (T=0) &=& \frac{\Delta^2_2}{8\pi} \int_0^\Lambda d \varepsilon \frac{\varepsilon}{((\varepsilon - \mu_2)^2 + \Delta^2_2)^{3/2}}.
    \label{nn_2_n}
    \eea
    The two stiffnesses are again related to the number of fermions in each band via
    $N_1 = 8 \pi N_0 \rho_{s,1}, N_2 = 8 \pi N_0 \rho_{s,2}$. Accordingly,
     \beq
    \rho_{comb} =  \rho_{s,1} + \rho_{s,2} = \frac{E_F}{4\pi}
     \label{nn_1_n}
     \eeq
      At $E_F \gg E_0$, $\rho_{comb} \gg T_{ins}$, hence $T_c \approx T_{ins}$. The two individual stiffnesses are
   \bea
  &&\rho_{e,2} (T=0) \approx 0.03 \Delta =0.05 T_{ins} \ll E_F\nonumber \\
  &&\rho_{s,1} (T=0) \approx \frac{E_F}{4\pi}.
   \eea
 This result is essentially identical to the one for the model with a hole and an electron pocket in the same limit.

 For $E_F = E_0$,  we obtain from Eq.(\ref{nn_2_n}), using the results for $\Delta_{e,h}$ and $\mu_{e,h}$ from Sec. \ref{sec_3nc},
 \bea
 \rho_{s,1} (T=0) & = &  0.053 E_F, \nonumber\\
 \rho_{s,2} (T=0) & = & 0.027 E_F,  \nonumber\\
 \rho_{comb}(T=0) & = &  0.08 E_F \equiv \frac{E_F}{4\pi}
 \eea
    Treating $1/(4\pi)$ as a small parameter and using the same estimate of the actual  $T_c$ as before we obtain $T_c \approx 0.125 E_F$.  The onset temperature for the pair formation is $T_{ins} = 0.924 E_F \approx  7.4 T_c$. Hence the system again displays preformed pair behavior  in a wide range of temperatures.  .

 Finally, for $E_F \ll E_0$, both $\mu_{1}$ and $\mu_2$ tend to $- E_0$, the gap behaves as $\Delta^2_1 = \Delta^2_2 = 2 E_F E_0$. Substituting
  into Eq.(\ref{nn_2_n}) we obtain
 \beq
 \rho_{s,1} \approx \rho_{s,2} = \frac{E_F}{8\pi}, ~\rho_{comb} = \frac{E_F}{4\pi}
 \label{ch_e11_n}
 \eeq
Hence $T_c \sim E_F$  and is parametrically smaller than $T_{ins} \sim E_0/\log{(E_0/E_F)}$, i.e., there is a parametrically wide range of preformed pair behavior.
This behavior is quite similar to the one in the canonical BEC regime, but we caution that $T_{ins}$ is still smaller by a large logarithm than the temperature $T_0\sim E_0$ at which a bound state of two fermions emerges in a vacuum.

\subsection{The density of states at $T=0$}

\begin{figure}[t!]
\renewcommand{\baselinestretch}{1.0}
$\begin{array}{c}
\includegraphics[angle=0,width=1.0\linewidth]{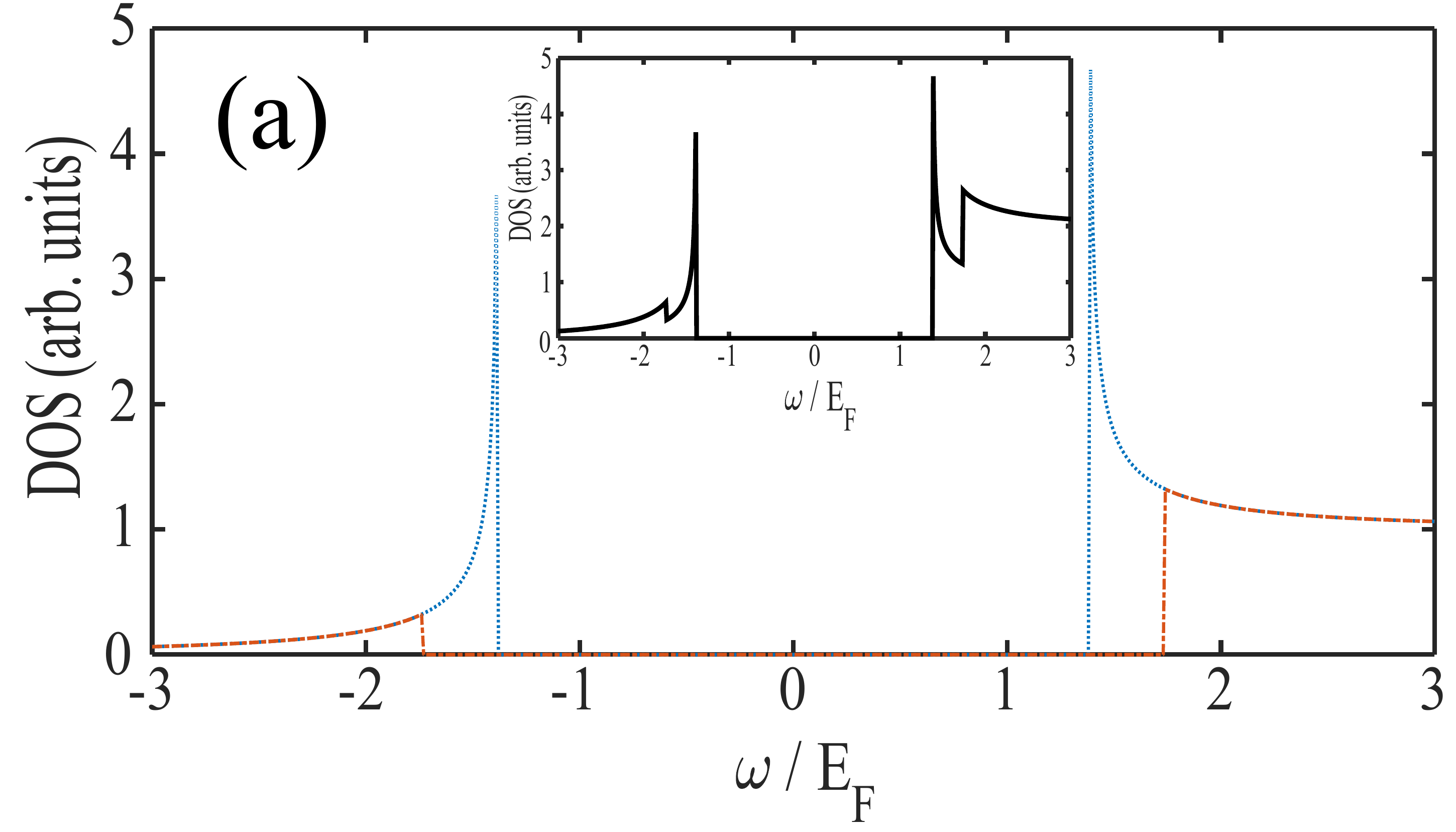} \\
\includegraphics[angle=0,width=1.0\linewidth]{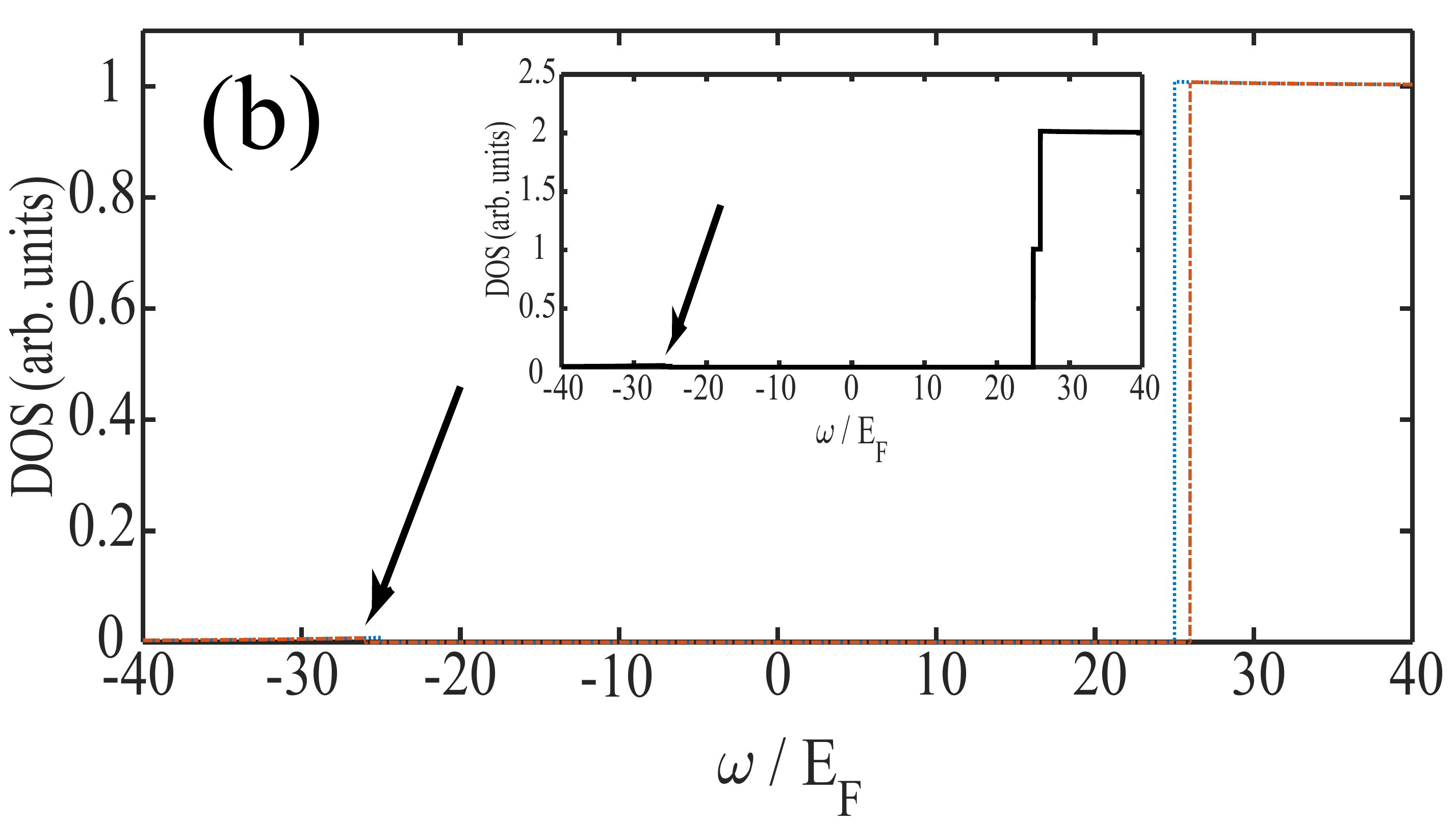}
\end{array}$
\caption{The DOS at $T=0$ for the model with two electron bands.
 Panel (a) -- $E_0 = E_F$,  panel (b) -- $E_0 =10 E_B$.
  For $E_0 = E_F$, $\mu_1 = -0.05 E_F, \mu_2 = -1.05 E_F$, and $\Delta_1 \approx - \Delta_e = 1.38 E_F$.  For $E_F=0.1E_0$,  $\mu_1 \approx -24.6 E_F$, $\mu_2 \approx -25.6 E_F$, and $\Delta_1 \approx \Delta_2 = 4.47 E_F$. We introduced the damping $\gamma=0.001E_F$ to make all features of the DOS visible. Main figures -- the DOS for each band (dashed red and blue lines).  Insets -- the total DOS.
  In the clean limit, the DOS has a discontinuity at $|\omega| =\sqrt{\Delta_i^2+\mu_i^2} +0$ ($i =1,2$), like in the one-band model. At $E_0 = E_F$, DOS on band 1 is large immediately after the jump because $\mu_1$ is very small.  The DOS for each band is  anisotropic between negative and positive frequencies due to anisotropy of the coherence factors.
 At large frequencies, the DOS tends to a finite value for positive $\omega$ and scales as $1/\omega^2$ for negative $\omega$. The arrow in panel (b) indicates the position of the would be discontinuity at a negative frequency. }
\label{fig_twobandee}
\end{figure}

The DOS at $T=0$ in the model with two electron bands is quite similar to that in the one-band model. Namely, the DOS is highly anisotropic and is much larger at positive frequencies than at negative frequencies.  For $E_F \ll E_0$, the chemical potentials are large and negative on both bands, and the DOS on each band is zero at $|\omega| < \sqrt{\mu^2 + \Delta^2}$, displays a discontinuity at $|\omega| = \sqrt{\mu^2 + \Delta^2}+0$,
 and at large negative frequencies scales as $1/\omega^2$ (see Eq. (\ref{nn_a})).
 We show the DOS for this model for $E_F= E_0$ and $E_F = 0.1 E_0$ in Fig. \ref{fig_twobandee}

\subsection{Evolution of $T_{ins}$ and $T_c$ with the filing of the second band and comparison with the experiments on Nb-doped SrTiO$_3$}
\label{sec_new_n}

Like we said in the Introduction,  superconductivity in the model with two electron bands is realized experimentally in Nb-doped SrTiO$_3$ and, possibly, in heterostructures of LaAlO$_3$ and SrTiO$_3$ (see Ref.\cite{Fernandes2013} and references therein).
   The Fermi energy in the band 1 is finite already at zero doping, and $E_F$  is likely larger than $E_0$, in which case $T_c \approx T_{ins}$.
     The band 2 is above the chemical potential at zero doping, but the chemical potential at $T=0$  moves up with doping and enters the band 2 once it
         exceeds the critical value.

    The experiments have found that superconducting $T_c$
 rapidly increases when the chemical potential enters the band 2.  This has been detected in Nb-doped SrTiO$_3$ (Ref. [\onlinecite{marel}])
  and in LaAlO$_3$/SrTiO$_3$ heterostructures (Ref. [\onlinecite{joshua}]).

 To verify whether this effect can be explained within our theory, we  extend our approach to the case when $E_F$ is finite in both bands.  To make notations more convenient, in this subsection we use $\mu$ instead of $E_F$ (see Fig. 3.)  We count $\mu$ from the bottom of the band 1, hence  the bare chemical potential $\mu_0$ (the one at $T=0$ and $\Delta =0$)
  coincides with $E_F$ in this band.   We assume that at $T=0$ the chemical potential in the band 2 is $\mu_0 - \mu^*$.  At zero doping $\mu^* > \mu_0$. At a finite doping, $\mu_0$ increases and crosses $\mu^*$ at some finite doping. Once $\mu_0$ gets larger than $\mu^*$,  the chemical potential  enters the band 2.
   We set $\mu_0 = \mu^* + \epsilon$, assume that $|\epsilon| \ll \mu_0$, and obtain the correction to $T_{c}$ to first order in $\epsilon$.

\begin{figure}[t!]
\renewcommand{\baselinestretch}{1.0}
\includegraphics[angle=0,width=1.0\linewidth]{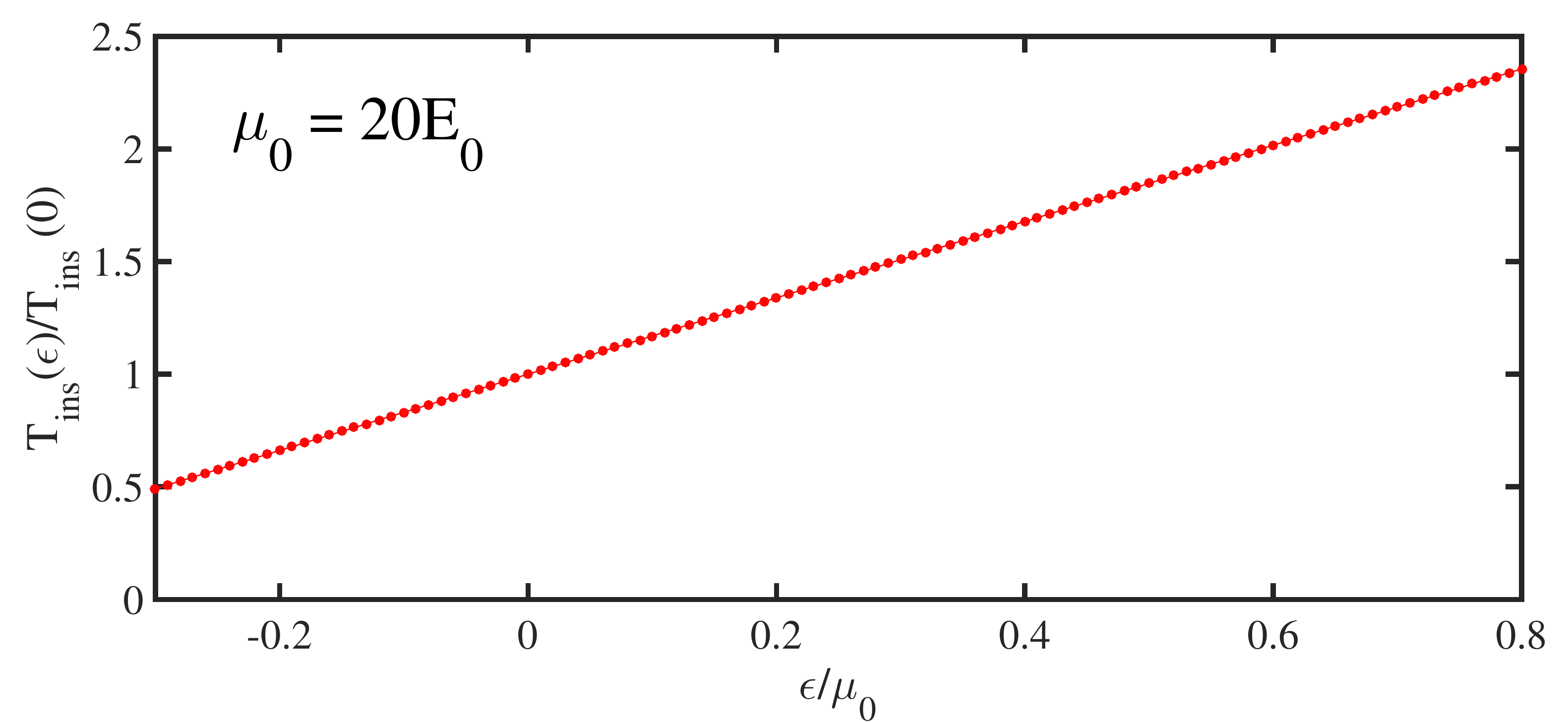}
\caption{Two-band model with two electron pockets. The slope of $T_{ins}$ vs $\epsilon/\mu_0$, where
$\mu_0 >0$ is the chemical potential in one of the bands and $\epsilon$
  is the deviation of $\mu_0$  from the bottom of one of the second band. Positive $\epsilon$ correspond to chemical potential crossing this second band.
   The results shown are for BCS limit $\mu_0 =20 E_0$, in which case $T_c \approx T_{ins}$.
 Observe that $T_{ins} \approx T_c$ rapidly increases when $\epsilon$ becomes positive. }%
\label{fig9}
\end{figure}

At a finite temperature, the chemical potentials in the two bands satisfy $\mu_1 = \mu_0 - \epsilon + \mu_2$ and the condition that the total number of particles is conserved reads
\beq
\mu_0 + \epsilon = T \log{\left[(1 + e^{\mu_{1}/T}) \times (1 + e^{\mu_{2}/T})\right]}
 \label{zzz_1}
 \eeq
 The onset temperature for the pairing, $T_{ins} \approx T_c$ is obtained by solving the set of linearized gap equations for $\Delta_1$ and $\Delta_2$.
 The set has the same form as in Eq. (\ref{ch_11_n}):
 \bea
\Delta_1 &=& - \frac{\lambda}{2} \Delta_2 \int_{-\mu_2}^{\Lambda}  \frac{dx}{x} \tanh{\frac{x}{2T_{ins}}} \nonumber \\
\Delta_2 &=& -  \frac{\lambda}{2} \Delta_1 \int_{-\mu_1}^{\Lambda}  \frac{dx}{x} \tanh{\frac{x}{2T_{ins}}}
\label{ch_11_nn}
\eea
 We assume that $\mu_0 \gg E_0$.  The analysis in Sec. (\ref{sec_3na}) for $\mu^* = \mu_0$ shows that $T_{ins} \ll \mu_0$, and, by continuity, we assume that this remains true for $\mu^* \approx \mu_0$.  One easily make sure that for such $T_{ins}$,  $\mu_1 (T_{ins}) \approx \mu_0 \gg T_{ins}$.  Eq. (\ref{zzz_1}) then reduces to
 \beq
 \mu_2 - 2 \epsilon = - T_{ins} \log{(1 + e^{\mu_{2}/T_{ins}})}
\eeq
Solving this equation we find
\beq
\mu_2 = - T_{ins} \log{\frac{2}{\sqrt{5} -1}} + \frac{4 \epsilon}{\sqrt{5}(\sqrt{5}-1)}
\eeq
Solving then the  linearized gap equation to first order in $\epsilon$  we obtain after a simple algebra
\beq
T_{ins} (\epsilon) = T_{ins} (0) + \frac{2\epsilon}{3\sqrt{5}(\sqrt{5}-1)}
\eeq
 where $T_{ins} (0) = 1.04 E^{2/3}_0 \mu^{1/3}_0$.
 Re-expressing this result back in terms of $\mu_0$ and $\mu^*$ we obtain
 \beq
T_{ins} (\epsilon) = T_{ins} (0) \left[1 + 0.23 \frac{\mu_0 - \mu^*}{\mu_0}  \left(\frac{\mu_0}{E_0}\right)^{2/3}\right]
\label{zzz_2}
\eeq
 We see that $T_{ins}$ and hence $T_c \approx T_{ins}$ increases once $\mu_0$ gets larger than $\mu^*$, and the slope is controlled by the large factor $(\mu_0/E_0)^{2/3}$, i.e., the relative
 increase is parametrically large.  By the same reason, $T_{c}$ rapidly decreases when $\mu_0$ is smaller than $\mu$.   We show this behavior in Fig. \ref{fig9}.

 That $T_c$ increases once $\mu_0$ gets larger than $\mu^*$ has been earlier found numerically in BCS calculations in Ref. [\onlinecite{Fernandes2013,Innocenti2010}].
   Our results are consistent with this work, however we emphasize that (i) our analytical result, Eq. (\ref{zzz_2}) shows that the slope of $T_c$ vs $\mu_0-\mu^*$ is parametrically enhanced at $\mu_0 \gg E_0$ and (ii) "pure" BCS calculation neglects the thermal evolution of the chemical potential, while in our analysis this renormalization is included and plays an important role.

 \section{The GMB  formalism}
 \label{sec_2a}

 A somewhat different approach to superconductivity in a situation when $E_F$ is much smaller than the upper energy cutoff for the pairing interaction was put forward by GMB in Ref. [\onlinecite{Gorkov61}].  They considered  weak coupling 3D case and argued that, from the physics point of view, $T_c$ has to be expressed in terms of observable quantum-mechanical scattering amplitude taken in the limit of zero momentum (the scattering length, $a$) rather than in terms of unobservable interaction potential $U$.  The relation between $U$ with $a$ is obtained by solving Schr\"odinger equation for one particle in a field $U$ of another particle. Diagrammatically, this amounts to summing up ladder series vertex correction diagrams in the particle-particle channel for two particles in a vacuum, i.e., for  zero chemical potential. To first order in $U$,  $a = mU/(4\pi)$
   and the  dimensionless coupling constant $\lambda = m|U|/(2\pi^2)$ (in 3D) equals to $(2 |a|k_F)/\pi$.  However, beyond leading order,
 $\lambda$ and $(2 |a|k_F)/\pi$ are not equivalent.

  GMB have demonstrated that, with logarithmic accuracy, the equation for $T_c$ in terms of $a$
 is obtained by summing up the same ladder series as in the BCS theory, however, two modifications have to be made simultaneously: (i) $U$ has to be replaced by $4\pi a/m$, and (ii) in each ladder cross-section one has to subtract from the product of the two Green's functions $G_{k,\omega} G_{-k, -\omega}$  the same $GG$  term taken at zero chemical potential.   As the result of these modifications, the kernel in the gap equation is cut at energies of order $E_F$, and $E_F$ appears as a prefactor in the formula for $T_c$, once the exponent contains $\pi/(2 |a|k_F)$ instead of $1/\lambda$.  GMB went further than logarithmic approximation and  obtained the exact weak coupling formula $T_c =0.277 E_F e^{-\pi/(2 |a|k_F)}$ by adding the leading renormalizations from the particle-hole channel.  We discuss these renormalizations in the Appendix.

 The GMB approach does not include phase fluctuations and hence the instability temperature obtained in this approach is actually the onset temperature for
 the pairing, $T_{ins}$.
 Below we discuss the extension of GMB approach to 2D case and show how our $T_{ins}$ can be re-expressed in terms of the physical scattering amplitude.
   We consider one-band model and two-band model with one hole and one electron pocket.  The analysis of the model with two electron pockets is equivalent to that of the one-band model. We show that GMB approach is applicable for arbitrary ratio of $E_F/E_0$, including the regime where $T_{ins}$ is  close to the temperature $T_0$, at which the scattering amplitude diverges.

 To set the stage, we first briefly review GMB approach in 3D and then consider 2D cases.

 \subsection{Original GMB consideration, weak coupling $D=3$ case}

To keep presentation short, we only restrict with the ladder series and neglect contributions from the particle-hole channel, i.e., will not try to reproduce the exact prefactor for $T_{ins}$.

GMB argued that to properly express $T_{ins}$ in terms of observable variables, one has to consider simultaneously the ladder series for the pairing vertex $\Phi$ in terms of
  bare $\Phi_0$ and ladder series for the vertex function $\Gamma = 4 \pi |a|/m$  for two particles in a vacuum in terms of the interaction $U$.  The ladder series are easily obtained diagrammatically and reduce to
 \begin{widetext}
  \bea
  \Phi &=& \Phi_0 \left[1 + \Pi_{pp} (E_F) |U| + \left(\Pi_{pp} (E_F) U\right)^2 + ...\right] = \frac{\Phi_0}{1-|U| \Pi_{pp} (E_F)} \nonumber \\
  \Gamma &=& |U| \left[1 + \Pi_{pp} (0) |U| + \left(\Pi_{pp} (0) U\right)^2 + ...\right] = \frac{|U|}{1-|U| \Pi_{pp} (0)},
  \label{s_1}
  \eea
  where
  \beq
  \Pi_{pp} (\mu) = T \sum_{\omega_m} \int d^3k/(2\pi^3) G (k, \omega) G(-k, - \omega) = T \sum_{\omega_m} \int d^3k/(2\pi^3) (\omega^2_m + (\varepsilon_k - \mu)^2)^{-1}.
  \eeq
  \end{widetext}

  Expressing $|U|$ in terms of $\Gamma$ as $|U| = \Gamma/(1 + \Gamma \Pi_{pp} (0))$ and substituting into the expression for $\Phi$ we obtain
  \beq
  \Phi =  \frac{\Phi^*_0}{ 1 - \Gamma \left(\Pi_{pp} (E_F) - \Pi_{pp} (0)\right)}
  \label{s_2}
  \eeq
  where $\Phi^*_0 = \Phi_0 (1 + \Gamma \Pi_{pp} (0)) \sim \Phi_0$.
  Eq. (\ref{s_2})  can be viewed as the sum of ladder series for $\Phi$ with $|U|$ replaced by $\Gamma = 4 \pi |a|/m$ and the term with zero chemical potential subtracted from $\int GG$.

  With these modifications, the equation for the instability temperature in terms of $|a|$  becomes:
\beq
1 = \frac{2 |a|k_F}{\pi}  \int_0^\Lambda d \varepsilon \left(\frac{\varepsilon}{E_F}\right)^{1/2} \left[\frac{\tanh\frac{\varepsilon - E_F}{2T_c}}{\varepsilon-E_F} - \frac{\tanh\frac{\varepsilon}{2T_c}}{\varepsilon}\right]
\label{ch_3}
\eeq
 We remind that we consider the case $E_F \ll \Lambda$.
 One can easily check that the integral over $\varepsilon$ now converges at $\varepsilon \sim E_F$, so the upper limit of integration doesn't matter any longer.
 The evaluation of the integral yields, at $|a| k_F \ll 1$,
 \beq
 T_c = 0.61 E_F e^{-\frac{\pi}{2 |a| k_F}}
 \label{ch_4}
 \eeq
 This equation re-expresses $T_{ins}$ in 3D in terms of the fully renormalized s-wave scattering length.
   The renormalizations from the particle-hole channel further change the prefactor to $0.277$ (Ref. \cite{Gorkov61}).

One can easily check that Eq. (\ref{ch_4})  coincides with the conventional BCS result for $T_{ins}$ in 3D.  Indeed, from the first equation in (\ref{s_1}) we obtain $T_{ins}$ in 3D directly in terms of $U$:
\beq
 T_{ins} = {\tilde \Lambda} e^{-\frac{1}{{\tilde \lambda}}}
\label{ch_5}
\eeq
where ${\tilde \lambda} = = m|U| k_F/(2\pi^2)$ (dimensionless coupling constant in 3D) and ${\tilde \Lambda} = 0.61 E_F e^{\sqrt{\frac{\Lambda}{E_F}}}$.
Using the  weak-coupling relation between $ak_F$ and ${\tilde \lambda}$:
 \beq
 \frac{2 |a| k_F}{\pi} \approx {\tilde \lambda} \left(1 + {\tilde \lambda} \sqrt{\frac{\Lambda}{E_F}}\right)
\label{ch_6}
\eeq
  one can easily verify that Eqs. (\ref{ch_4}) and (\ref{ch_5}) are equivalent, as they indeed should be.

We emphasize that, although Eqs. (\ref{ch_4}) and (\ref{ch_5}) are identical, the physics behind GMB approach in 3D is the separation of scales: fermions with energies below $E_F$ are the only ones which contribute to superconductivity, while fermions with energies above $E_F$  renormalize  the interaction between low-energy fermions into the quantum-mechanical scattering amplitude.

\subsection{Extension of GMB formalism to 2D}
\label{sec_2b}

We now extend this approach to 2D case.  We consider separately one-band and two-band models.
  \subsubsection{One-band model}

   We first consider the case $E_F > E_0$, when bound pairs do not develop prior to superconductivity and the scattering amplitude
 is small at weak coupling, and then extend the analysis to the case $E_F < E_0$.

 In the 2D case the scattering amplitude, which we label $a_2$, is dimensionless.  To first order in $U<0$ we still have $a_2 = mU/(4\pi)$.
   Keeping $a_2$ as a small parameter
 and performing GMB computation of $T_{ins}$ in 2D,  we obtain
 \beq
 T_{ins} = 1.13 E_F e^{-\frac{1}{|a_2|}}
 \label{ch_7}
 \eeq
This equation is the 2D analog of Eq. (\ref{ch_4}). It expresses $T_{ins}$ in terms of the 2D scattering amplitude, which is an observable variable.

Eq. (\ref{ch_7}) is the same as Eq. (\ref{s_3}) for $T_{ins}$ in terms of $U$, as we now demonstrate.
 Summing up ladder series of renormalizations which convert  $U$ into $m a_2/(4\pi)$, we obtain
\beq
\frac{1}{a_2} =   \log{\frac{1.13 \Lambda}{T}} - \frac{2}{\lambda}
\label{ch_8}
\eeq
 where, we remind, $\lambda = m|U|/(2\pi)$.   At  $E_F > E_0$, $T_{ins} \sim (E_F E_0)^{1/2} \gg E_0$, hence at $a_2$ is negative at $T \sim T_{ins}$,  like the interaction  $U$.
Substituting $1/|a_2| = -1/a_2$ from Eq.(\ref{ch_8}) into Eq.(\ref{ch_7}) we obtain
\beq
T_{ins} = 1.13 (\Lambda E_F)^{1/2}  e^{-\frac{1}{\lambda}}
\label{ch_9}
\eeq
 This coincides with  Eq. (\ref{s_3}).

 In the opposite limit $E_F \ll E_0$,  the temperature $T_{ins}$ is larger than $E_F$, hence the temperature dependence of the chemical potential $\mu$ must be included into the GMB-type analysis.  Performing the same calculation as before and treating $a_2$ as some temperature-dependent parameter, not necessary a small one, we obtain, using $\mu = -T \log{T/E_F}$:
 \beq
 \log{\log{\frac{T_{ins}}{E_F}}} = \frac{1}{a_2}
 \label{fr_1}
 \eeq
 or
 \beq
 T_{ins} = E_F e^{e^{1/a_2}}
 \label{fr_2}
 \eeq
 This formula again expresses $T_{ins}$ in terms of the scattering amplitude $a$, with $E_F$ as the overall factor.

 Eq. (\ref{fr_2}) looks simple, but one should keep in mind that the scattering amplitude $a_2$ by itself depends on temperature. In view of this, $T_{ins}$ is actually the solution of the transcendental equation $T = E_F \exp{(\exp{(1/a_2 (T))})}$, in which $a_2(T)$ should be treated as  input function, extracted from independent measurements.

 We now demonstrate that, although $T_{ins}$ in Eq.(\ref{fr_2}) contains $E_F \ll E_0$ as the overall factor, this $T_{ins}$  coincides with  that in Eq.(\ref{s_4}), once $a_2$ is re-expressed
 back in terms of $\lambda$.  To see this we substitute $1/a_2$ from Eq.(\ref{ch_8}) into Eq.(\ref{fr_1}) and obtain
 \beq
 \log{\frac{T_{ins}}{E_F}} =  e^{-\frac{2}{\lambda}} e^{\log{\frac{1.13\Lambda}{T_{ins}}}} = \frac{E_0}{T_{ins}}
 \eeq
 or
 \beq
  T_{ins} = \frac{E_0}{\log{\frac{E_0}{E_F}}}
 \label{fr_3}
 \eeq
 This is the same expression as Eq.(\ref{s_4}) as it indeed should be.

 Note that $a_2$ is actually positive at $T = T_{ins}$ because $T_{ins}$ is smaller than $T_0=1.13 E_0$  at which a bound state forms in a vacuum.
     Taken at a face value, this would reply that the interaction becomes repulsive. However, one can easily verify that $a_2(T_{ins})$ changes sign
exactly when $\mu (T_{ins})$ crosses zero. As a result, $\int (GG(\mu) - GG(0))$ becomes negative simultaneously with the sign change of $a$ and the product
$a_2 *\int (GG(\mu) - GG(0))$  in the denominator of (\ref{s_2}) remains positive.

 The case when $T_{ins} \approx T_0$ actually requires more sophisticated treatment because the scattering amplitude is large at $T = T_{ins}$  and the corrections to the ladder diagrams, which we neglected, may become relevant. We will not pursue this case nor discuss the mathematical details how to properly extend the ladder series
  for the scattering amplitude in Eq.(\ref{ch_8}) to the case when $a$ in Eq.(\ref{ch_8}) changes sign.  We just consider the agreement between Eqs. (\ref{fr_2}) and (\ref{s_4}) is the evidence that Eq. (\ref{ch_8}) can be used even when $a_2$ is not small.

\subsection{Two-band model}

We now extend GMB analysis to the two band model  with a hole and an electron bands.

 The inter-band scattering amplitude $a_{he}$ is again obtained by summing up  ladder diagrams for the vertex function for the two particles in a vacuum. Inter-band scattering is reproduced in  odd orders in the interaction.  We set  $m_h=m_e$ to simplify calculations, sum up odd terms in the ladder series,  and obtain
   \beq
  a_{he} =  \frac{\lambda}{2} \frac{1}{1-\frac{\lambda^2}{4} \Pi^2}
  \eeq
  where $\Pi =  \log{1.13 \Lambda/T}$ is the particle-particle polarization operator at $\mu =0$. We remind that $\lambda = m|U|/(2\pi)$ in 2D.

  Expressing $\lambda$ in terms of $a_{he}$ as
  \beq
  \lambda = \frac{4 a_{he}}{1+ \sqrt{1 + 4 a^2_{he} \Pi^2}}
  \eeq
  and substituting this back into the set of linearized equations for $\Delta_1$ and $\Delta_e$,  Eqs. (\ref{ch_11}), we obtain after a simple
  algebra the equation for $T_{ins}$ as a function of $a_{he}$:
   \beq
   2 + 2 \sqrt{1 + 4 a^2_{he} \Pi^2} = 4 a^2_{he} \left(\Pi_e \Pi_h - \Pi^2\right)
  \label{fri_1}
   \eeq
    where
    \bea
    \Pi_e &=& \int_{-\mu_e}^\Lambda d\varepsilon \frac{\tanh{\frac{\varepsilon}{2T}}}{\varepsilon} \nonumber \\
    \Pi_h &=& \int_{-\mu_e}^\Lambda d\varepsilon \frac{\tanh{\frac{\varepsilon}{2T}}}{\varepsilon}
    \label{fri_2}
    \eea
    The r.h.s. of Eq.(\ref{fri_1}) can be re-written as
    \beq
    \left(\Pi_e \Pi_h - \Pi^2\right) = {\tilde \Pi}_e {\tilde \Pi}_h + \Pi \left({\tilde \Pi}_e {\tilde \Pi}_h \right)
    \eeq
     and
     \bea
     {\tilde \Pi}_{e} &=& \Pi_e - \Pi = \int_{0}^{\mu_e} d\varepsilon \frac{\tanh{\frac{\varepsilon}{2T}}}{\varepsilon} \nonumber \\
      {\tilde \Pi}_{h} &=& \Pi_h - \Pi = \int_{0}^{\mu_h} d\varepsilon \frac{\tanh{\frac{\varepsilon}{2T}}}{\varepsilon} \nonumber \\
   \label{fri_3}
      \eea
 Each ${\tilde \Pi}$ is $\int (GG (\mu) - GG (0))$, i.e., it is the difference between the particle-particle polarization operator at a finite chemical potential $\mu_{e,h}$ and the one at $\mu=0$.   The chemical potentials $\mu_{e,h}$ are at most of order $E_F$, hence the integrals in  Eq.(\ref{fri_3}) come from energies below $E_F$, like in the original GMB analysis.

 At small $E_0/E_F$, $T_{ins}$ is smaller than $E_F$, $\mu_e =  - 0.48 T_{ins}$ and $\mu_e \approx E_F  \gg T_{ins}$ (see Eq.(\ref{fri_4})).
 Then  ${\tilde \Pi}_e = O(1)$, while ${\tilde \Pi}_h \approx \log{(1.13 E_F/T_{ins})} \gg 1$
 In this situation,  $\Pi_e \Pi_h - \Pi^2  \approx \Pi {\tilde \Pi}_h$.  Substituting this into Eq. (\ref{fri_1}) we obtain that, up to an overall factor,
 \beq
 T_{ins} = E_F e^{- \frac{1 + \sqrt{1 + a^2_{12} \Pi^2}}{2 a^2_{he} \Pi}}
 \label{fri_5}
 \eeq
   This is the transcendental equation on $T_{ins}$ with temperature dependence in the r.h.s. coming from $\Pi = \log{1.13 \Lambda/T_{ins}}$ and
    from $a_{he} = a_{he} (T)$.  The latter again should be treated as input parameter, extracted from independent measurements.

   It is  straightforward to verify that $T_{ins}$ in Eq.(\ref{fri_5}) is the same as we obtained in Eq. (\ref{ch_12}) earlier
    in terms of the coupling $\lambda$ (or, equivalently, in terms of $E_0$).  To see this, we re-express $a_{he}$ back in terms of $\lambda$. This converts Eq. (\ref{fri_5}) into
    \beq
    T_{ins} = E_F e^{-\frac{4 \left(1 - \frac{\lambda^2 \Pi^2}{4}\right)}{\lambda^2 \Pi}} = E_F e^{-\frac{4}{\lambda^2 \Pi}} \frac{\Lambda}{T_{ins}}
    \eeq
    hence, up to constant prefactors,
    \beq
    \log{\frac{(\Lambda E_F)^{1/2}}{T}} \log{ \frac{\Lambda}{T}} = \frac{2}{\lambda^2}
    \eeq
    Using $\log{\Lambda} = (2/\lambda) + \log{E_0}$ we obtain after simple algebra, $T_{ins} \sim E^{1/3}_F E^{2/3}_0$, what agrees with Eq.(\ref{ch_12}).

    In the opposite limit $E_0 > E_F$, we use the fact that $T_{ins} \gg E_F$, $\mu_e \approx -E_F/2$, $\mu_h \approx 3 E_F/2$, and obtain ${\tilde \Pi}_e \approx - E_F/(4T_{ins})$ and ${\tilde \Pi}_h \approx 3E_F/(4T_{ins})$. Substituting into  Eq.(\ref{fri_1}) and using the fact that $\Pi \gg 1$, while ${\tilde \Pi}_{e,h}$ are small, we obtain
    \beq
    T_{ins} = \frac{E_F}{2} \frac{2 a^2_{he} \Pi}{1 + \sqrt{1 + 4 a^2_{he} \Pi^2}}
    \label{fri_6}
    \eeq
    Using next the fact that at small $E_F$, $T_{ins}$ is close to $T_0$ at which the scattering amplitude diverges, i.e, $a_{he} \Pi \gg 1$, we can further approximate Eq.(\ref{fri_6}) to
    \beq
    T_{ins} =  E_F \frac{a_{he}}{2}
    \label{fri_7}
    \eeq
    Note that in the two-band case $a_{he}$ remains positive (like $U$) for arbitrary $E_F/E_0$  because even at $E_F \to 0$, $T_{ins}$ is larger than $T_0$, see (\ref {ch_16}).

  Eq. (\ref{fri_7}) expresses $T_{ins}$  at $E_F \ll E_0$ in terms of $E_F$ and 2D inter-band scattering amplitude, which, again, should be considered as temperature dependent input function.

  One can easily demonstrate that Eq. (\ref{fri_7}) coincides with Eq.(\ref{ch_16}). For this we note that, when $a_{he}$ is large, it can be expressed via the coupling $\lambda$ (or, equivalently, via $E_0$) as
  \beq
  a_{he} \approx \frac{1}{2\log b},
  \eeq
  where $b \log b \approx (E_F/4 E_0)$.  For small $E_F/E_0$, $\log{b} \approx E_F/4 E_0$, hence $a_{he} \approx 2E_0/E_F$.  Substituting this into Eq. (\ref{fri_7}), we obtain
  $T_{ins} \approx E_0$. This coincides with Eq.(\ref{ch_16}) up to corrections which we neglected by approximating $a_{he}$ by $2 E_0/E_F$.

\section{Conclusions}
\label{sec_4}

In this paper we considered the interplay between superconductivity and formation of bound pairs of fermions  in multi-band 2D fermionic systems (BCS-BEC  crossover). In two spatial dimensions a  bound state develops already at weak coupling, and BCS-BEC  crossover can be analyzed already at weak coupling, when calculations are fully under control.
  We reviewed the situation in one-band model and considered two different two-band models, one with one hole and one electron band and the other with
  two hole or two electron bands  The first model is relevant to experiments on Fe-pnictides and Fe-chalcogenides, particularly on FeSe,
   the second one is used to describe Nb-doped SrTiO$_3$.

     For each model we solved self-consistently the equations  for the gaps and the chemical potentials on the two bands and  obtained the onset temperature of the pairing, $T_{ins}$,  and the chemical potentials and the pairing gaps below $T_{ins}$.
      We computed superfluid stiffnesses and used  them to estimate the actual superconducting $T_c$ below which $U(1)$ gauge symmetry is spontaneously broken.

    In a one-band model,  the system displays BCS behavior when the Fermi energy $E_F$  exceeds the energy, $E_0$, of a bound state of two fermions in a vacuum.
    In this regime, (i) $T_{ins} \sim (E_F E_0)^{1/2}$  is parameterically larger than  $E_0$, i.e., the pairing emerges at a much higher $T$ than would-be the temperature for the bound state formation in a vacuum, and (ii) the superfluid stiffness $\rho_s = E_F/(4\pi)$ is parametically larger than $T_{ins}$, hence phase fluctuations are costly  near $T_{ins}$. As the consequence, phase coherence sets in almost immediately after bound pairs form.   In the opposite limit $E_0 \gg E_F$, the pairing develops at $T_{ins} \gg E_F$, while the actual $T_c$ is determined by phase fluctuations and is of order $E_F$.  In between $T_{ins}$ and $T_c$ bound pairs develop but remain incoherent, and the fermionic spectral function displays a pseudogap behavior, when the spectral weight gradually transforms from the Fermi surface to an energy of order of the pairing gap $\Delta \sim (E_F E_0)^{1/2}$.  This is  a typical system  behavior in the BEC regime.  The only difference with the "canonical" BEC behavior in 3D, where strong coupling is a must, is that in our weak coupling model  bound pairs are not tightly bound molecules because the two fermions in a pair are separated on average at distances by a distance well above the interatomic spacing.  We argud that the fermionic spectral function is highly non-symmetric in the preformed pairs regime.

We next considered the two-band model with one hole and one electron band.
       For definiteness we set $E_F =0$ on the electron band, but kept $E_F$ finite on the hole band.
        We found that the behavior of this model is different in several aspects from that in the one-band model.
       There is again a crossover from BCS-like behavior at $E_F \gg E_0$ to BEC-like behavior at $E_F \ll E_0$ with $T_{ins} > T_c$.
        However,  in distinction to the one-band case,
          the actual $T_c$, below which long-range superconducting order develops, remains finite and of order $T_{ins}$ even when $E_F=0$ on both bands.  The reason for a finite $T_c$ is that the filled hole band acts as a reservoir of fermions. The pairing reconstructs fermionic dispersion and transforms some spectral weight into the
          newly created hole band below the original electron band and electron band above the original hole band.  A finite density of fermions in these two bands gives rise to a finite $T_c$  even when the bare Fermi level is exactly at the bottom of the electron band and at the top of the  hole band.

  We also considered the model with two hole/two electron bands. We found that the behavior in this model is similar to that in the one-band model.  Namely, BCS-BEC crossover occurs when the largest of the two $E_F$'s  becomes comparable to $E_0$.  When the ratio $E_F/E_0$ is large, the system displays BCS-like behavior, when it is  small, the system displays the same BEC-type behavior as in the one-band model, namely $T_c$ scales with $E_F$ and is parametrically smaller than $T_{ins}$.

Finally, we re-expressed $T_{ins}$ in terms of the 2D scattering amplitude, which is a physical observable, in distinction to $U$.  For this, we extended to $D=2$ the approach put forward by Gorkov and Melik Barkhudarov back in 1961 for $D=3$ case.   We obtained the explicit formulas for $T_{ins}$ in terms of the 2D dimensionless
scattering amplitude $a_2$  for the one-band model and for the model with one hole and one electron band, and demonstrated that these formulas are valid not only in the BCS limit but also in the BEC limit, when the scattering amplitude is not small.   One distinction between 2D and 3D cases is that in our 2D case the scattering amplitude $a_2(T)$ is temperature dependent, hence the formulas relating $T_{ins}$ and $a_2(T_{ins})$ are transcendental equations, which have to be solved with $a_2(T)$ taken from a separate measurement.

Our results confirm earlier BCS analysis by several groups~\cite{Chen2015,phillips2015} that in the one hole/one electron band model $T_{c}$ doesn't tend to zero if $E_F$ on one band vanishes, and it remains finite even when one of the bands is located entirely below or entirely above the Fermi level.  However,
 previous works  identified $T_{ins}$ with $T_c$, while we show that $T_c \approx T_{ins}$ only when $E_F \gg E_0$, while at $E_F < E_0$
    $T_c$ is numerically substantially smaller than $T_{ins}$.
    Our results for $T_{ins}$ also differ from these earlier works because they
     neglected the temperature dependence of the chemical potential.

 With respect to applications to Fe-based superconductors, our results do confirm that $T_c$ does not vanish when one of hole bands sinks below the Fermi level ot moves as a whole above the Fermi level. Furthermore, the gap on this band is generally higher than that on the bands which cross the Fermi level.  The gap ratio is non-universal and depends on the mass ratio and/or presence of additional bands. That the gap is larger on the band that does not cross the Fermi level is consistent with the experimental results reported for LiFe$_{1-x}$Co$_x$As in
  Ref. \cite{Miao2015} and for FeTe$_{0.6}$Se$_{0.4}$ in Ref. \cite{Ozaki2014}. We note in passing that
   one does not need to invoke  a highly unconventional concept of the ultra-strong pairing at all momenta in the Brillouin zone~\cite{Miao2015} to explain the data.

 Our analysis for the case $E_F \leq E_0$ may be relevant to FeSe.  In this material Fermi energies on all bands are only a few $meV$, and are comparable to $T_c$.
  For the two-band model, we found that $T_{ins}$  and $T_c$ do differ by a sizable factor, and there exists an intermediate $T$ range of preformed pair behavior.
    Recent experiments on FeSe have been interpreted~\cite{shibauchi} in terms of pre-formed pairs which appear at about twice $T_c$.  This is exciting possibility and the theoretical study of the interplay between $T_{ins}$ and $T_c$ in the full multi-band model for FeSe is clearly called for.

Finally, our analysis of  the model with two electron pockets one of which has a finite $E_F$ and for the other the Fermi level is near its bottom,  may be relevant to superconductivity in  Nb-doped SrTiO$_3$
  (Refs. \cite{Fernandes2013,marel,Innocenti2010,Bianconi2013}) and LaAlO$_3$/SrTiO$_3$ heterostructures (Ref. [\onlinecite{joshua}]). These materials contain two electron bands,  and the Fermi level passes through the bottom of one of the bands upon doping.  Experiments have found~\cite{marel,joshua} that $T_c$ rapidly increases once the chemical potential moves up and crosses both bands.
   We reproduced this result in our theory -- we found that $T_c$ increases when the chemical potential moves into the  second band, and the slope of the increase of $T_c$ contains a large parameter.

 We acknowledge with thanks the discussions with  C. Batista, S. Borisenko, A. Charnukha, L.P. Gorkov, M. Grilli, R. Fernandes, P. Hirschfeld, C.A.R. Sa de Melo, S. Maiti, D. Mozursky, Y. Matsuda,  I. Mazin,  T. Shibauchi, A. Varlamov,  and, particularly, with L. Benfatto and M. Randeria.   We thank L. Benfatto for pointing out an error in the earlier version of the manuscript. The work was supported by the Office of Basic Energy Sciences U. S. Department of Energy
under award DE-SC0014402 (AVC). The work of IE was supported by the Focus Program 1458 Eisen-Pniktide of the DFG and by the German Academic
Exchange Service (DAAD PPP USA no. 57051534).  AVC acknowledges with thanks support by LANL through  Ulam fellowship.

\section{Appendix}

In this Appendix we obtain  the exact  prefactor for $T_{ins}$ in Eqs. (\ref{s_3}) and (\ref{s_4}).
 To compute it, one needs to go beyond ladder approximation and include the fermionic self-energy to order $\lambda$ and  the corrections to $U$ from the particle-hole channel.

\begin{figure}[t!]
\renewcommand{\baselinestretch}{.8}
\includegraphics[angle=0,width=1.0\linewidth]{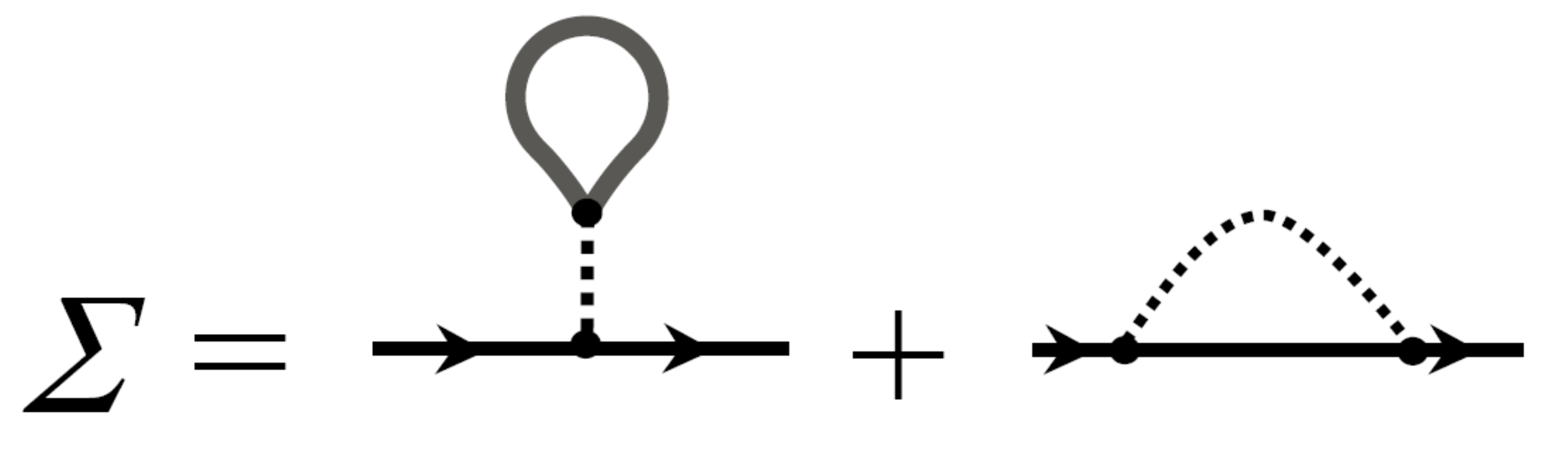}
\caption{Hartree-Fock diagrams for the self-energy.}
\label{fig6}
\end{figure}

 The fermionic self-energy to order $\lambda$ comes from Hartree and Fock diagrams in Fig. \ref{fig6}.  This self-energy renormalizes fermionic dispersion and the chemical potential, and
   also changes fermionic residue to $Z <1$. The correction to $Z$ originates from frequency-dependent part of the fermionic self-energy $\Sigma (k, \omega)$. The latter is  non-zero in our model, despite that the interaction is approximated by the static $U$, because we set the sharp frequency cutoff at energy scale  $\Lambda$ (and momentum cutoff at $\varepsilon_k = \Lambda$).

\begin{figure}[t!]
\renewcommand{\baselinestretch}{.8}
\includegraphics[angle=0,width=1.0\linewidth]{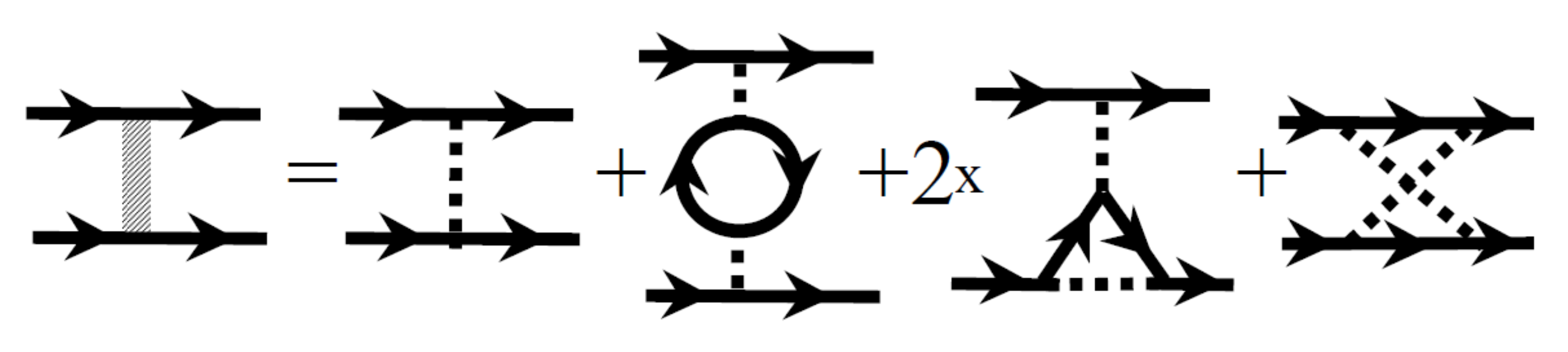}
\caption{Second order diagrams for the  renormalization of the irreducible pairing interaction due to contributions from the particle-hole channel.}
\label{fig7}
\end{figure}

 We assume that the renormalization of $\mu$ at $T=0$  is already incorporated into $E_F$. The  remaining one-loop self-energy  has the form
 \beq
 \Sigma (k, \omega) = i \omega \frac{\lambda}{4} f_\omega\left(\frac{\omega}{\Lambda}, \frac{\varepsilon_k}{\Lambda}\right) - \varepsilon_k  \frac{\lambda}{2\pi} f_\varepsilon \left(\frac{\omega}{\Lambda}, \frac{\varepsilon_k}{\Lambda}\right),
 \label{ch_e2}
 \eeq
  where the scaling functions satisfy $f_\omega(0,0) = f_\varepsilon (0,0) =1$. We define the sign of the self-energy via $G^{-1} (k, \omega) = i\omega - \varepsilon_k + \Sigma (k, \omega)$.
   The self-energy comes from internal frequency and momentum of order of the upper cutoff and only weakly depends on $E_F \ll \Lambda$.  The scaling functions $f_\omega$ and $f_\varepsilon$ can be straightforwardly obtained numerically. However, for our purposes we will need the renormalization of the fermionic propagator only at $\omega \sim \varepsilon_k \sim T_{ins} \ll \Lambda$.  At these energies the scaling functions can be approximated by their values at $\omega = \varepsilon_k =0$.  The full Green function to order $\lambda$ in this energy/momentum range is then
   \beq
   G(k,\omega) = \frac{Z}{\omega - \left(\frac{k^2}{2m^*} - E_F \right)},
  \label{ch_e3}
   \eeq
   where
   \beq
   Z = 1 - \frac{\lambda}{4},~~ m^* = m \left(1 + \frac{\lambda}{4} - \frac{\lambda}{2\pi}\right)
   \label{ch_e4}
   \eeq

Substituting the Green's function from (\ref{ch_e3}) into  the ladder diagrams, we immediately obtain that the fermionic self-energy changes the dimensionless coupling $\lambda$
  into
  \beq
  {\tilde \lambda} = \lambda m^* Z^2 = \lambda \left( 1 - \lambda \frac{\pi +2}{4\pi}\right)
  \label{ch_e5}
  \eeq

  The renormalization of the irreducible pairing interaction to order $\lambda$ comes from
   particle-hole channel and generally involves four diagrams, each contains a particle-hole bubble (Fig. \ref{fig7}). For a constant  interaction $U$  the first three diagrams in Fig. 6 cancel out and only the last, exchange diagram contributes.  In 2D the contribution from this diagram to the irreducible coupling (i.e., the correction to $U$)
    is  a constant, equal to $ U \lambda$,  when relevant transferred frequency is much smaller than $E_F$. At a higher frequency $\Omega$ the renormalization from the exchange diagram is additionally reduced by $E_F/\Omega$.
      Accordingly, the renormalization of $U$ by the particle-hole bubble is only relevant at $E_F \gg E_0$, when  $T_{ins} \ll E_F$.
    In this regime, the effective coupling  constant is
    \beq
 \lambda_{eff} =  {\tilde \lambda} (1 - \lambda)  = \lambda \left( 1 - \lambda \frac{5\pi +2}{4\pi}\right)
  \label{ch_e5a}
  \eeq
  In the opposite limit,  $E_F \ll E_0$, $T_{ins} \gg E_F$, and the renormalization from particle-hole channel can be neglected.
   In this regime, the effective coupling is
      \beq
 \lambda_{eff} =  {\tilde \lambda}  = \lambda \left( 1 - \lambda \frac{\pi +2}{4\pi}\right)
  \label{ch_e5b}
  \eeq
    Collecting all renormalizations to order $\lambda$ and substituting into Eqs. (\ref{s_3}) and (\ref{s_4}), we obtain for $T_{ins}$ at $E_F \gg E_0$,
    \beq
 T_{ins}  =  1.13 (\Lambda E_F)^{1/2} e^{-1/{\tilde \lambda}} = 0.276 (\Lambda E_F)^{1/2}  e^{-1/\lambda}
 \label{s_3a}
 \eeq
 The prefactor differs somewhat from that in Ref. [\onlinecite{Gorkov61}b].
At $E_F  \ll  E_0$ we have
  \beq
  T_{ins}  = 1.13 \frac{\Lambda}{\log{E_0/E_F}}  e^{-2/{\tilde {\tilde \lambda}}} = 0.751
  \frac{E_0}{\log{E_0/E_F}}
  \label{s_4a}
  \eeq


\begin{thebibliography}{100}
\bibitem{mohit} The authors are thankful to M. Randeria for clarifying this issue.
\end{thebibliography}
\end{document}